\documentclass[5p,times]{elsarticle}
\usepackage{amsmath}
\usepackage{graphicx}
\usepackage{amsmath,amssymb,amsfonts}
\usepackage{pgf-pie}
\usepackage{textcomp}
\usepackage{longtable}
\usepackage{xcolor,color}
\usepackage{xcolor}
\usepackage{xpatch}
\usepackage{subfig} 
\usepackage{enumerate,enumitem}
\usepackage{algorithm}
\usepackage{algorithmic}
\usepackage{mathtools}
\usepackage{blkarray}
\usepackage{lscape}
\usepackage{multirow}
\usepackage{lipsum}
\usepackage{pifont}%
\newcommand{\cmark}{\ding{51}}%
\newcommand{\xmark}{\ding{55}}%
\usepackage{fancyhdr}
\usepackage{color, colortbl}

\usepackage{url}

\usepackage{todonotes}
\usepackage[capitalise]{cleveref}

\journal{}

\begin{document}
\begin{frontmatter}

\title{Towards Message Brokers for Generative AI: Survey, Challenges, and Opportunities}

\author{Alaa~Saleh}
\ead{alaa.saleh@oulu.fi}
\author{Susanna~Pirttikangas}
\ead{susanna.pirttikangas@oulu.fi}
\author{Lauri~Lovén\corref{cor1}}
\ead{lauri.loven@oulu.fi}
\address{are with the Center for Ubiquitous Computing, University of Oulu, Finland.}
\cortext[cor1]{Corresponding author}
\author{Roberto Morabito}
\ead{roberto.morabito@eurecom.fr}
\address{is with the Department of
Communication Systems, EURECOM, France.}
\author{Sasu Tarkoma}
\ead{sasu.tarkoma@helsinki.fi}
\address{is with the Department of
Computer Science, University of Helsinki, Finland.}

\begin{abstract}
In today's digital world, Generative Artificial Intelligence (GenAI) such as Large Language Models (LLMs) is becoming increasingly prevalent, extending its reach across diverse applications. This surge in adoption has sparked a significant increase in demand for data-centric GenAI models, highlighting the necessity for robust data communication infrastructures. Central to this need are message brokers, which serve as essential channels for data transfer within various system components. This survey aims to delve into a comprehensive analysis of traditional and modern message brokers, offering a comparative study of prevalent platforms. Our study considers numerous criteria including, but not limited to, open-source availability, integrated monitoring tools, message prioritization mechanisms, capabilities for parallel processing, reliability, distribution and clustering functionalities, authentication processes, data persistence strategies, fault tolerance, and scalability. Furthermore, we explore the intrinsic constraints that the design and operation of each message broker might impose, recognizing that these limitations are crucial in understanding their real-world applicability. Finally, this study examines the enhancement of message broker mechanisms specifically for GenAI contexts, emphasizing the criticality of developing a versatile message broker framework. Such a framework would be poised for quick adaptation, catering to the dynamic and growing demands of GenAI in the foreseeable future. Through this dual-pronged approach, we intend to contribute a foundational compendium that can guide future innovations and infrastructural advancements in the realm of GenAI data communication.
\end{abstract}


\end{frontmatter}

\section{Introduction}


In the burgeoning field of Generative Artificial Intelligence (GenAI), the computing continuum faces unprecedented challenges in efficiently managing data flows and computational resources. GenAI has predominantly targeted consumer applications, offering applications such as the ChatGPT \cite{chatgpt}, a conversational AI agent based on a large language model (LLMs), a type of machine learning (ML) model with a deep neural network. However, a surge in machine-to-machine (M2M) use cases, coupled with the increasing possibility of relying more and more on decentralized, distributed, and edge-based Large Language Models (LLMs), beckons a reevaluation of the supporting communication infrastructure. This reevaluation is necessitated by the evolving demands for higher bandwidth, lower latency, and more robust data processing capabilities that these advanced applications require \cite{shen2024large, tarkoma2023ai}.
This survey paper delves into the role of publish/subscribe (pub/sub) message broker systems, considering in particular their emerging role for seamless and scalable data exchange in GenAI applications. We scrutinize contemporary message brokers for their adaptability and efficiency in GenAI contexts, outline existing challenges, and chart promising research avenues for future development.



In more detail, while the focus of GenAI systems has been on consumer-oriented applications, there is an increased interest towards M2M use cases. GenAI has been proposed to be used in, for example, for networking, wireless communication, and compression~\cite{bariah2023large}.

As a result, current computing continuum platforms, spanning the networks and computational resources from user devices to cloud~\cite{kokkonen2022autonomy,motlagh2022edge}, face new challenges. These platforms provide support AI models, offering interconnect between their data sources and sinks, and optimising their use of resources in the computing continuum. As GenAI models require and generate ever larger amounts of data~\cite{wang2023overview}, all the while consuming computational resources varying from moderate to massive~\cite{kokkonen2022autonomy}, the computing continuum must offer a dynamic and scalable communication and computation substrate to ensure timely data dissemination and efficient use of resources~\cite{dustdar2022distributed,Praveen2023exploring}.

Pub/sub approach is equally useful alongside other paradigms across the computing continuum for a wide range of AI applications, including smart cities~\cite{8782370}, healthcare~\cite{7098706}, and many other domains. This approach decouples data producers from their consumers, allowing applications to develop components independently, and enhancing system robustness and adaptability~\cite{eugster2003many}. Furthermore, pub/sub makes system design more flexible by increasing the independence between system components with a reliable interconnect.

Pub/sub systems are based on the exchange of data between clients (e.g., services or application components) through a message broker. Publishers submit content to the broker, which then allows subscribers to access that content without knowing its source~\cite{tarkoma2012publish,donta2022survey}. By managing, filtering, and routing communication between publishers and subscribers, the message broker acts as an intermediary layer~\cite{hasenburg2020managing}, routing and distributing messages efficiently, accurately, and in a timely manner, based on the interests expressed by subscribers. Within the brokers, message queues can temporarily store the messages, protecting the system from overflows or outages. Moreover, brokers also often provide other essential features such as persistent storage, monitoring, and authentication.




As GenAI continues to evolve, parallels can be drawn with the historical trajectory of IoT systems, where the emergence of robust and adaptive message brokers marked a significant evolutionary step. These message brokers became a de facto paradigm, primarily because they addressed critical challenges associated with scalability, real-time data processing, and the integration of heterogeneous devices and platforms \cite{redondi2019towards}. Similarly, in the context of GenAI, the anticipation of an analogous development is not without merit. The complexity and volume of data that GenAI applications demand, coupled with the necessity for high-quality service monitoring and data exchange processes, suggest that a transition towards more sophisticated message brokering solutions may be inevitable. Such solutions would not only have to manage the large data throughput but also ensure adaptability, reliability, and efficiency in dynamic GenAI ecosystems. Reflecting on the IoT evolution, the motivations for this shift include the need to support scalable communication, facilitate interoperability among diverse systems, and uphold stringent Quality of Service (QoS) standards, which are likely to be paralleled in the GenAI domain ~\cite{loven2023can}.

However, it is critical to acknowledge that the evolution towards more sophisticated message brokering solutions, specifically tailored to accommodate GenAI application needs, does not come without its challenges.

Bearing all this in mind, the primary contributions of this survey are summarized as follows:

\begin{itemize}
    \item We provide a comprehensive review of recent message brokers, evaluating the brokers by  their suitability for GenAI systems. We aim to guide the development of a compatible brokering framework in consideration of the evolving requirements for GenAI systems in the future.
    \item We summarize the challenges of message brokers and highlight the need for a robust and efficient data communications substrate based on an increase in GenAI applications.
    \item We discuss central research topics and their potential focus areas for making message brokers suitable for GenAI applications. We also describe promising algorithms for implementing such brokers.

\end{itemize}

This paper's remaining sections are organized as follows. Section~\ref{sec:Pubsubparadigm} provides a general definition of pub/sub paradigm and highlights the advantages and disadvantages of both broker-based and brokerless messaging architectures. Section~\ref{sec:studyonMQ} presents existing message brokers with their features and cons. Section~\ref{sec:ways} examines possible ways to make message brokers suitable for GenAI applications. The paper concludes with Section~\ref{sec:conclusion}.

\section{The Pub/Sub Paradigm}\label{sec:Pubsubparadigm}

Pub/sub is a messaging paradigm where publishers send messages without indicating specific recipients. Remaining oblivious to the original publishers, subscribers receive relevant messages according to their interests. At its core, pub/sub thus decouples message delivery from the senders and recipients. This enhances the system's adaptability and robustness, as it allows publishers and subscribers to operate independently~\cite{eugster2003many}. Furthermore, subscribers can flexibly choose topics based on their interests, enabling them to find content relevant to their preferences. As a result of pub/sub, real-time messaging can be sent to a wide range of subscribers, enabling scalable and timely dissemination of information~\cite{donta2022survey,chafi2022introduction,maniezzo2023self}. 

As part of the pub/sub communication model, a message broker functions as an intermediary layer, managing the flow of messages from publishers to subscribers. By ensuring that messages are accurately routed to subscribers, based on their expressed interests or specific topics, the broker ensures that messages are received by subscribers precisely as they have been requested. By providing a layer of abstraction between publishers and subscribers, the broker goes beyond simply facilitating message transmission. Therefore, neither party needs to be aware of the other's operations or presence. A key strength of the broker is its reliability~\cite{pedrosa2021reducing}, as it has mechanisms for guaranteeing delivery of messages even when subscribers are temporarily offline or have connectivity difficulties. 

\begin{figure}[t]
\centering
\includegraphics[width=0.5\textwidth]{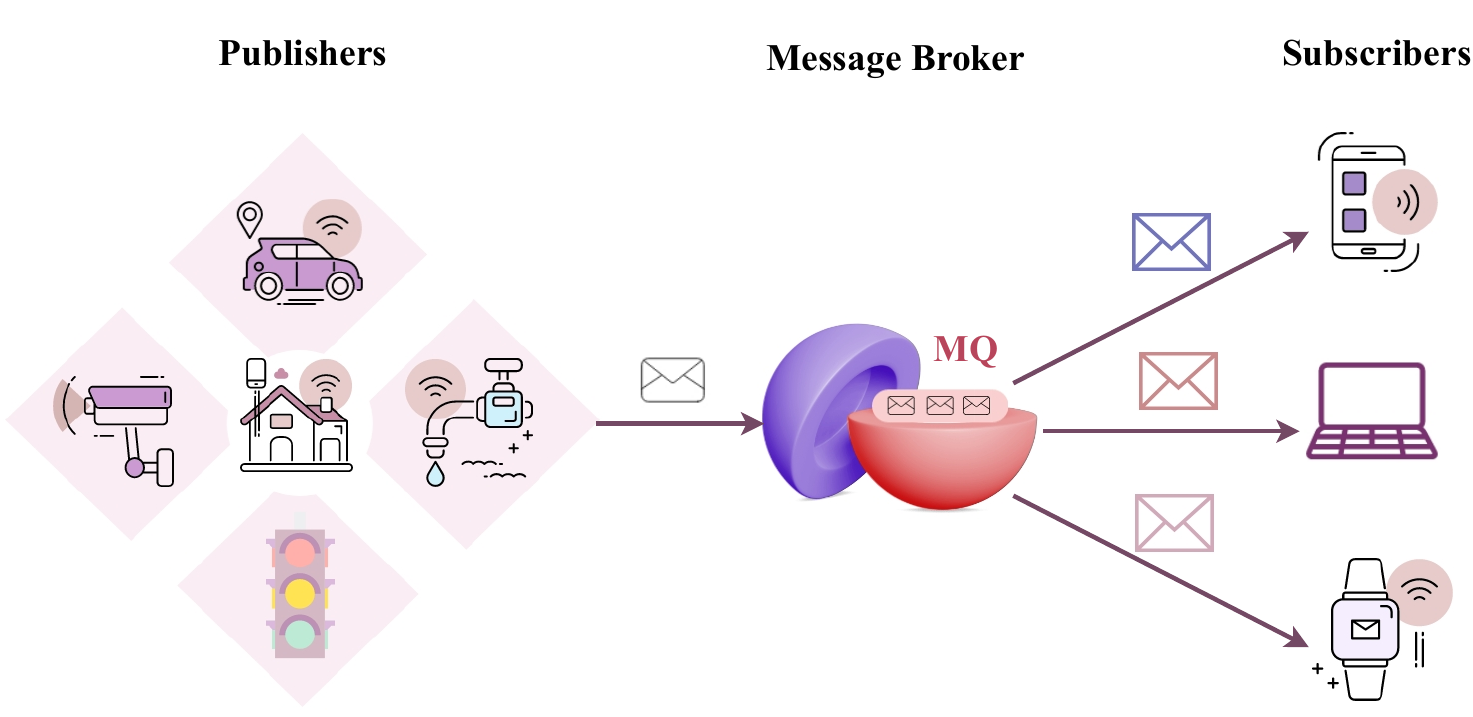}
\caption{The Publish/Subscribe paradigm.}\label{fig:broker}
\end{figure}

Moreover, message queues are general-purpose components of the broker that temporarily store messages from publishers until they can be delivered to subscribers as part of the broker process~\cite{john2017survey}. By orchestrating the overall flow of messages, the message broker ensures that the appropriate subscribers receive the messages based on their subscriptions, while a message queue ensures that these messages are held and dispatched in an orderly manner, ensuring that publishers and subscribers are able to communicate in an asynchronously and decoupled ways as shown in \cref{fig:broker}. 


\subsection{Message broker development}

\begin{figure}[t]
\centering
\includegraphics[width=0.5\textwidth]{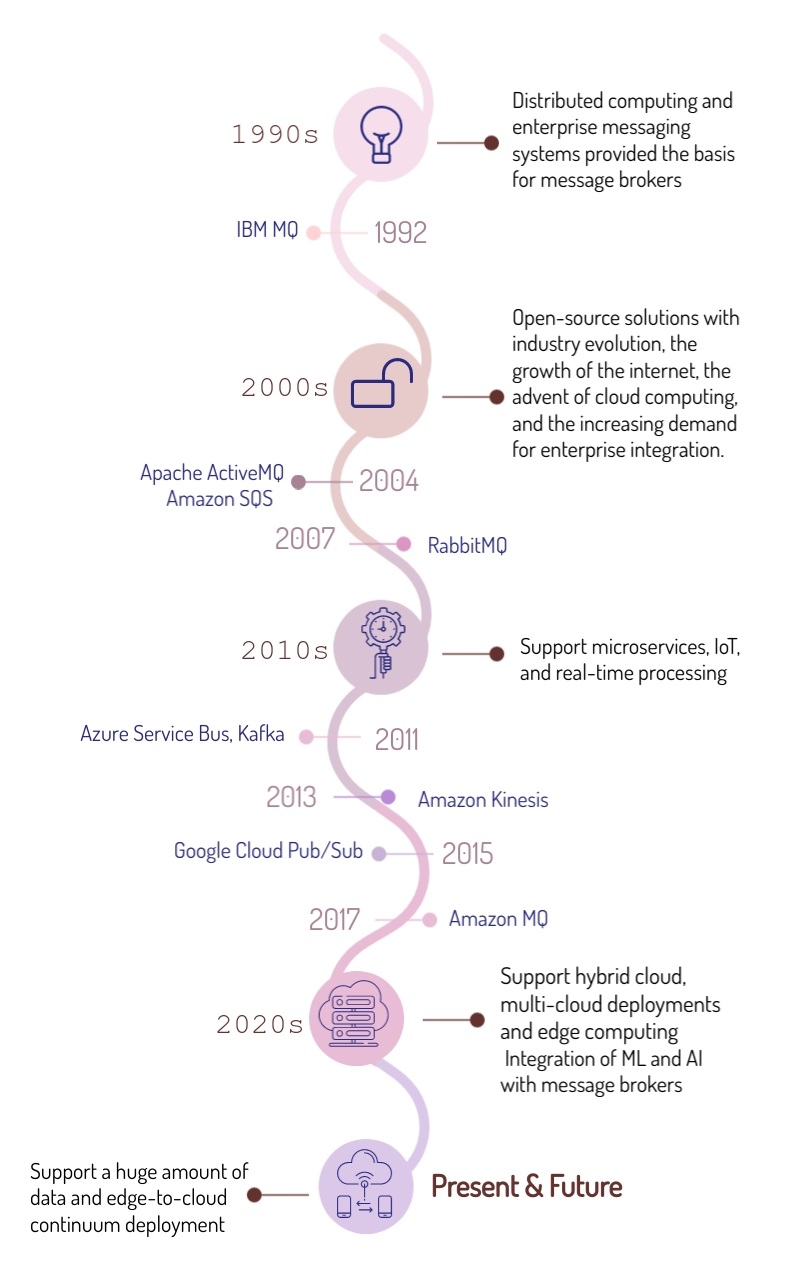}
\caption{The timeline of message broker evolution from 1990 to present.}\label{fig:timeline}
\end{figure}

Over the past 30 years, message broker technology has evolved and innovated significantly as shown in Fig.~\ref{fig:timeline}. With the development of message-oriented middleware (MOM) and the Java Message Service (JMS), the technology began to gain prominence between 1980 and 1999, as organizations began to require greater integration and communication~\cite{linthicum2000enterprise,perry2001mqseries}. During the period 2000-2009, the technology underwent significant advancements, influenced by the implementation of service-oriented architecture (SOA), the increased internet usage, and the emergence of cloud computing~\cite{hohpe2004enterprise}. Among these developments was the implementation of web services standards and open-source alternatives, laying the foundation for future advances in cloud-based message brokers. During the following decade, from 2010 to 2019, the technology adapted to new demands such as real-time data processing, cloud computing, IoT, and microservices architectures~\cite{john2017survey}. Increasing demands for Internet of Things and real-time data processing have led to the rise of containerization platforms such as Docker and Kubernetes.

With the advent of cloud-native architectures, microservices, edge computing, and IoT demands, message broker technology continued to evolve unabated between 2020 and 2023~\cite{indrasiri2021design}. The integration of AI and ML was particularly important for optimizing message routing, anomaly detection, and auto-scaling, addressing the complexities of growing data volumes. In the future, message broker technology will continue to evolve, leveraging advances in computing and communication. As 5G/6G technologies advance, cross-platform interoperability, and decentralized architectures are developed, a number of trends are set to shape its future. These include edge computing, quantum computing, enhanced security, serverless architectures, and advancements in the 5G/6G technologies. 

\subsection{Broker vs. Brokerless Messaging Architecture}
A broker is the main unit for managing and monitoring data of pub/sub systems \cite{tarkoma2012publish}, offering scalability, balanced load distribution, and optimal resource utilization, among others. Additionally, the message broker ensures that messages are reliably transmitted, preventing any loss of data. 


Despite their benefits, certain challenges accompany message brokers, particularly concerning scalability and efficiency. As data volume grows, brokers face increased complexity, undermining scalability -- a common issue also observed in brokerless pub/sub systems known for their simplicity, quick access, and improved efficiency. In such systems, publishers and subscribers interact directly, making discovery, management, and availability crucial factors. However, the absence of a central unit for overseeing message flows complicates supervision and control. Furthermore, this setup does not inherently guarantee reliable message delivery~\cite{iakushkin2014messaging}.

\section{Survey of Message Brokers}\label{sec:studyonMQ}

In the past decade, numerous message brokers have been developed both in proprietary and open-source sectors. Each broker possesses unique features and pitfalls, influenced by their respective vendors and intended applications. This section focuses on the most commonly used message brokers, with their features and challenges concisely summarized in \cref{tab:MBs1,tab:MBs2,tab:MBs3,tab:MBs4}. In following subsections, we categorize these message brokers based on their open-source availability and the priority-based delivery of messages (built-in priority-support) that ensures messages are delivered in priority order, with high-priority messages processed first.


 
\subsection{Open Source Message Brokers}

We found 30 message brokers that were available as open source. Out of these, 17 supported priority messages, while 13 did not. Each is discussed in more detail in below subsections.

\subsubsection{Priority Support}
\textit{Apache ActiveMQ}~\cite{ApacheActiveMQ} is a Java-based message broker licensed under the Apache 2.0 license. Through the use of double layers of SSL/TLS security layers, it provides dual security levels. A distribution of Apache ActiveMQ provided by FuseSource, \textit{Fuse Message Broker}~\cite{Fuse} supports J2EE integration capabilities such as Java Database Connectivity (JDBC), J2EE Connector Architecture (JCA), and Enterprise JavaBeans (EJB). \textit{Apache Qpid}~\cite{ApacheQpid}, on the other hand, provides cloud-based messaging capabilities and supports queuing for structured message exchange, making it essential for distributed applications.

\textit{RabbitMQ}~\cite{RabbitMQ} was developed by Rabbit Technologies Ltd in 2006 using the Erlang programming language and released under the Mozilla Public License. It supports multiple protocols, including AMQP, STOMP, and MQTT. \textit{HornetQ}~\cite{HornetQ}, a Java application based on JBoss, provides a distributed messaging platform for enterprise-level applications using STOMP and AMQP protocols.

\textit{Red Hat AMQ}~\cite{RedHatAMQ} is a messaging protocol based on Java for large-scale Internet business applications with no administrative costs, installation, or configuration required. \textit{Celery}~\cite{Celery} is written in Python and supports multiple message brokers, including RabbitMQ and Redis. Using Java Messaging Service (JMS) API, \textit{JBoss Messaging}~\cite{JBossMessaging} is a messaging broker provided by JBoss, a division of Red Hat for facilitating communication between different components or applications in a distributed system.

\textit{OpenMQ}~\cite{OpenMQ} is implemented in Java and was developed by Oracle as an open source protocol. \textit{Beanstalk} \cite{Beanstalk} creates queues automatically with pure Python. \textit{Gearman}~\cite{Gearman} is an optimized server written in C/C++ with a simple interface that provides low application overhead. \textit{Enduro/X}~\cite{Enduro/X} is written in C and offers native APIs for C/C++. For enhanced interprocess communication, it utilizes in-memory POSIX kernel queues. As part of the WSO2 Integration platform, \textit{WSO2 Message Broker}~\cite{WSO2} is a message-based communication component.

\textit{HiveMQ}~\cite{HiveMQ} is compatible with MQTTv3.1 and all subsequent versions. The Eclipse Public License (EPL) and Eclipse Distribution License (EDL) cover this implementation. \textit{Redis}~\cite{Redis} is BSD-licensed, used by companies such as Uber, Instagram, and AirBNB for caching and messaging queues. With 100 million concurrent connections per cluster and sub-millisecond latency, \textit{EMQX}~\cite{EMQX} can efficiently and reliably connect massive amounts of IoT devices. EMQX nodes can be bridged by other MQTT servers and cloud services to send messages across platforms. Additionally, it deploys and operates on all public cloud platforms. \textit{Apache Pulsar}~\cite{Pulsar} is an open-source distributed messaging system developed as a queuing system, but it recently added event streaming features. It combines many Kafka and RabbitMQ features.

\subsubsection{No Priority Support}
\textit{Apache Kafka}~\cite{Kafka} was developed by LinkedIn as a distributed streaming platform, supporting multiple data formats, including JSON, Avro, and XML. Furthermore, Java, Python, and Go are the official client libraries and several cloud platforms are supported, including Amazon Web Services, Microsoft Azure, and Google Cloud Platform. It provides a variety of tools for managing and monitoring Kafka clusters, such as Kafka Manager, Kafka Monitor, and Kafka Connect. 

\textit{Apache RocketMQ}~\cite{ApacheRocketMQ} is a cloud-native platform that operates across distributed systems, facilitating real-time data processing. With support for versions 5.0, 3.1.1, and 3.1, \textit{Eclipse Mosquitto}~\cite{EclipseMosquitto} implements the MQTT protocol. \textit{ZeroMQ}~\cite{zeromq} is supported by a large and active open source community, and utilizes a broker-less pub/sub pattern. 

\textit{Apache NiFi}~\cite{ApacheNiFi}, developed by the Apache Software Foundation, automates data exchange between software systems, and facilitates the conversion of data formats in real time.  \textit{Ably Realtime}~\cite{AblyRealtime} is built on Ably's Data Stream Network, which includes a cloud network and realtime messaging fabric. Additionally, over 40 Client Library SDKs are available, as well as native support for six real-time protocols. \textit{Apache SamZa}~\cite{SamZa} is a streaming framework based on Apache Kafka and Apache Hadoop developed by LinkedIn and now part of the Apache Software Foundation. It processes real-time data streams generated by Apache Kafka, Amazon Kinesis, and Azure Event Hub.

 \textit{VerneMQ}~\cite{VerneMQ} was launched in 2014 by Erlio GmbH. It supports MQTT messages in LevelDB, and uses a clustering architecture based on Plumtree, however it isn't actively developed and lacks features. \textit{NServiceBus}~\cite{NServiceBus} is designed with simplicity. With a number of retry strategies, a message which fails processing can automatically be forwarded to an error queue for manual investigation. \textit{Kestrel}~\cite{kestrel} is a JVM-based distributed message queue inspired by Blaine Cook's "Starling". 

In \textit{NSQ}~\cite{NSQ}, distributed and decentralized topologies are promoted, allowing fault tolerance and high availability as well as reliable delivery by replicating every message across multiple nodes within the cluster. \textit{NATS}~\cite{NATS} was originally released in 2011 and was written in Go. \textit{KubeMQ}~\cite{KubeMQ} is a modern and innovative message queue and broker that facilitates communication across cloud platforms, on-premise environments, and edge deployments.

\subsection{Proprietary Message Brokers}

We found 16 proprietary message brokers, out of which 10 supported priority messages while 6 did not. Each is discussed in more detail in below subsections.

\subsubsection{Priority Support}
\textit{IBM MQ}~\cite{ibmmq} supports data exchange between applications, systems, services, and files via messaging queues, serving as a crucial communication layer for message flow management. It offers flexibility in deployment options, whether in virtual machines or containers, including Docker, Kubernetes/Cri-O and Red Hat OpenShift. Moreover, it is ideal for applications demanding high reliability and zero message loss. \textit{Amazon Simple Queue Service}~\cite{SQS} is operated by Amazon, so it can handle a lot of traffic with providing authentication using the Amazon API key and secret. However, requests are sent to the SQS web service via HTTP, which is susceptible to latency issues. 

\textit{Microsoft Message Queue}~\cite{MSMQ} is a messaging infrastructure created by Microsoft and built into the Windows Operating System. It serves as a queue manager and allows two or more applications to communicate without immediately knowing each other's responses. 

As a Java-based message broker, \textit{Oracle GlassFish Server Message Queue }~\cite{Oracle1} provides message brokering services to popular message queue systems such as AQ, IBM MQ Series, and TIBCO Rendezvous. It provides a consistent, open, JMS-compliant API for these message queuing systems. Additionally, OMB supports both durable and non-durable subscribers, as well as the JMS standard pub/sub, topic-based routing.  

\textit{TIBCO Rendezvous}~\cite{TIBCORendezvous} is a peer-to-peer architecture for high-speed data distribution. \textit{TIBCO Enterprise Message Service}~\cite{TIBCOEnterpriseMessageService} is a message oriented middleware that supports a wide range of message protocols and technologies, including the Java Message Service (JMS) standard using Java and J2EE, Microsoft .NET, TIBCO FTL, TIBCO Rendezvous and C and COBOL on the Mainframe. Besides supporting up to 10 MB payloads in XML, JSON, CSV, HTML and plain text formats, \textit{Anypoint MQ}~\cite{AnypointMQ} also has easy connectivity to Mule applications or non-Mule applications. 

\textit{Azure Service Bus}~\cite{AzureServiceBus} from Microsoft is a cloud-based message broker that only supports AMQP and STOMP protocols. As a fundamental part of \textit{SAP NW PI}~\cite{SAP} architecture, an Integration Broker facilitates communication between different enterprise applications, both SAP-based and non-SAP. 

\textit{Solace Message Broker}~\cite{Solace}, also known as Solace PubSub+, is an advanced event broker that facilitates the efficient exchange of information between applications, IoT devices, and users through several messaging paradigms, including pub/sub, queue, request/reply, and streaming.

\subsubsection{No Priority Support}

\textit{Google Cloud Pub/Sub}~\cite{GoogleCloudPub/Sub} is a messaging service offered by Google Cloud. In addition to native integration with other Google Cloud services, including Cloud Functions, Dataflow, and BigQuer, as well as a variety of development tools like Cloud Shell, Cloud Code, and Cloud Build, it supports real-time data processing for ML applications with Google Cloud AI Platform and other ML services. Aside from the Stackdriver Logging and Stackdriver Monitoring tools, it also provides SDKs for Java, Python, Node.js, and Go. 

\textit{Amazon MQ}~\cite{AmazonMQ} developed for ActiveMQ based on Java with support for MQTT, AMQP, STOMP, and WebSocket. \textit{Intel MPI Library}~\cite{IntelMPILibrary} provides a cloud support for Amazon Web Services, Microsoft Azure, and Google Cloud Platform. \textit{Amazon Kinesis}~\cite{Kinesis} is a real-time streaming data service with a scalable and durable architecture that can capture and store GBs or TBs of data per second from multiple sources for up to 24 hours. It provides various developer tools and integrations with AWS services, such as SDKs, templates, and integrations with AWS CloudFormation. 

\textit{Azure Storage Queue}~\cite{AzureStorageQueue} provides cloud messaging that enhances communication in the cloud, on desktops, on-premises, and on mobile devices. \textit{IronMQ}~\cite{IronMQ} runs on public clouds as well as on-premise with providing client libraries in a wide variety of programming languages, including Python, Ruby, Java, PHP, and NET. 

\begin{table*}[]
\centering
\caption{A summary of open source and priority-supporting message brokers.}
\label{tab:MBs1}
\resizebox{!}{.39\paperheight}{%
\begin{tabular}{|c|c|c|c|c|c|c|c|c|c|c|c|l|l|}
\hline
\rowcolor[HTML]{F4EFEF} 
\rotatebox{90}{\textbf{\begin{tabular}[c]{@{}c@{}}Message \\ Brokers and Queues\end{tabular}}} &
  \rotatebox{90}{\textbf{Clustering Support}} &
  \rotatebox{90}{\textbf{Monitoring Support}} &
  \rotatebox{90}{\textbf{Pub/Sub Support}} &
  \rotatebox{90}{\textbf{Parallel Processing}} &
  \rotatebox{90}{\textbf{Pull \& Push Support}} &
  \rotatebox{90}{\textbf{Reliable Delivery}} &
  \rotatebox{90}{\textbf{Persistent}} &
  \rotatebox{90}{\textbf{Authentication}} &
  \rotatebox{90}{\textbf{Scalable}} &
  \rotatebox{90}{\textbf{Distributed}} &
  \rotatebox{90}{\textbf{Fault Tolerance}} &
  \multicolumn{1}{c|}{\cellcolor[HTML]{F4EFEF}\rotatebox{90}{\textbf{Shortcomings}}} &
  \multicolumn{1}{c|}{\cellcolor[HTML]{F4EFEF}\rotatebox{90}{\textbf{Features}}} \\ \hline
\rowcolor[HTML]{FFFFFF} 
\textbf{Apache ActiveMQ~\cite{ApacheActiveMQ}} &
  \cmark &
  \cmark &
  \cmark &
  \cmark &
  Both &
  \cmark &
  \cmark &
  \cmark &
  \cmark &
  \cmark &
  \cmark &
  \begin{tabular}[c]{@{}l@{}}Message delivery guarantees are limited~\cite{ApacheActiveMQ}.\\ Memory per queue is limited, the default number \\of messages is 400~\cite{ApacheActiveMQ}.\\ Installation is complex~\cite{ApacheActiveMQ}.\\ Scaling is challenging~\cite{ApacheActiveMQ}.\end{tabular} &
  \begin{tabular}[c]{@{}l@{}}Multi-protocols \& multi-languages support~\cite{ApacheActiveMQ}.\\ Efficient management and resource allocation~\cite{ApacheActiveMQ}.\\ Supports flow control and message expiration~\cite{ApacheActiveMQ}.\\ Provides message groups as well as virtual and combined queues~\cite{ApacheActiveMQ}.\\ Works on small and medium-scale applications~\cite{ApacheActiveMQ}.\end{tabular} \\ \hline
\rowcolor[HTML]{F4EFEF} 
\textbf{Fuse Message Broker~\cite{Fuse}} &
  \cmark &
  \cmark &
  \cmark &
  \cmark &
  Both &
  \cmark &
  \cmark &
  \cmark &
  \cmark &
  \cmark &
  \cmark &
  \begin{tabular}[c]{@{}l@{}}Limited monitoring tools\cite{Fuse1}.\end{tabular} &
  \begin{tabular}[c]{@{}l@{}}Written in Java~\cite{Fuse1}.\\ Supports JMS 1.1 \& J2EE 1.4 integration-related components~\cite{Fuse1}.\\ Supports loosely couple applications~\cite{Fuse1}.\\ Supports multi-languages including C/C++, Java, .NET, Ruby, Perl, \\PHP, Pike,\& Python~\cite{Fuse1}. \\ Supports message compression~\cite{Fuse1}.\end{tabular} \\ \hline
\rowcolor[HTML]{FFFFFF} 
\textbf{Apache Qpid~\cite{ApacheQpid}} &
  \cmark &
  \cmark &
  \cmark &
  \cmark &
  Both &
  \cmark &
  \cmark &
  \cmark &
  \cmark &
  \cmark &
  \cmark &
  \begin{tabular}[c]{@{}l@{}}Compatibility issues between versions~\cite{ApacheQpid}.\\Message size is limited to 100MB for AMQP \\protocols 0-8, 0-9, or 0-91~\cite{ApacheQpid} \end{tabular} &
  \begin{tabular}[c]{@{}l@{}}Implements AMQP Protocol~\cite{ApacheQpid}. \\ Easy to use~\cite{ApacheQpid2}. \\ Detects failures and assigns messages to different brokers~\cite{ApacheQpid}.\\ Low latency~\cite{ApacheQpid}. \\ Supports multiple authentication schemes~\cite{ApacheQpid}.\\Active connections can be limited to protect client processes \\from malicious activity~\cite{ApacheQpid}.\end{tabular} \\ \hline
\rowcolor[HTML]{F4EFEF} 
\textbf{RabbitMQ~\cite{RabbitMQ}} &
  \cmark &
  \cmark &
  \cmark &
  \cmark &
  Push &
  \cmark &
  \cmark &
  \cmark &
  \cmark &
  \cmark &
  \cmark &
  \begin{tabular}[c]{@{}l@{}}Written in Erlang, which is unfamiliar to many \\developers~\cite{RabbitMQ}. \\ Queues with large numbers of messages are memory-\\intensive and strain brokers~\cite{RabbitMQ}.\\ Redundant message broker communication~\cite{donta2022survey}.\\ Clustering has few features and is complicated~\cite{RabbitMQ}.\\ Message size is limited to 512MB~\cite{donta2022survey,RabbitMQ}. \end{tabular} &
  \begin{tabular}[c]{@{}l@{}}Runs on all major operating systems~\cite{RabbitMQ}.\\ Has good documentation~\cite{RabbitMQ}.\\ Works with C, C++, .NET, and Python~\cite{RabbitMQ}. \\ Supports asynchronous cluster-to-cluster message routing~\cite{RabbitMQ1}. \\ Supports multiple messaging protocols~\cite{RabbitMQ}.\\ Offers several built-in exchange types~\cite{RabbitMQ}. \\ Supports flow control for balancing workloads and avoiding rapid \\messages flooding~\cite{RabbitMQ}.\end{tabular} \\ \hline
\rowcolor[HTML]{FFFFFF} 
\textbf{HornetQ~\cite{HornetQ}} &
  \cmark &
  \cmark &
  \cmark &
  \cmark &
  Both &
  \cmark &
  \cmark &
  \cmark &
  \cmark &
  \cmark &
  \cmark &
  \begin{tabular}[c]{@{}l@{}}Data loss may occur~\cite{donta2022survey}.\\ Delay may occur with large messages (up to 100KB) \\due to split message into multiple packages~\cite{HornetQuser}.\end{tabular} &
  \begin{tabular}[c]{@{}l@{}}Supports AMQP and STOMP protocols~\cite{HornetQprotocol}.\\ Provides better performance and stability when combined with \\ActiveMQ~\cite{donta2022survey}.\end{tabular} \\ \hline
\rowcolor[HTML]{F4EFEF} 
\textbf{Red Hat AMQ~\cite{RedHatAMQ}} &
  \cmark &
  \cmark &
  \cmark &
  \cmark &
  Both &
  \cmark &
  \cmark &
  \cmark &
  \cmark &
  \cmark &
  \cmark &
  \begin{tabular}[c]{@{}l@{}}Queue access is limited by special characters~\cite{RedHatAMQ1}. \\ Non-persistent messages are lost when brokers \\stop~\cite{RedHatAMQ1}.\end{tabular} &
  \begin{tabular}[c]{@{}l@{}}Enables real-time integration~\cite{RedHatAMQ}.\\ Supports multi-message patterns for real-time messaging~\cite{RedHatAMQ}.\\ Supports multi-languages, including Java, C, C++, Python, Ruby,\\\& .Net~\cite{RedHatAMQ}.\\ Supports mission-critical applications~\cite{RedHatAMQ}. \end{tabular} \\ \hline
\rowcolor[HTML]{FFFFFF} 
\textbf{Celery~\cite{Celery}} &
  \cmark &
  \cmark &
  \cmark &
  \cmark &
  Push &
  \cmark &
  \cmark &
  \xmark &
  \cmark &
  \cmark &
  \cmark &
  \begin{tabular}[c]{@{}l@{}}Compatibility and integration with other brokers \\can be complicated~\cite{Celery}.\\ Overall complexity~\cite{Celery}.\\ Monitoring and management are challenging~\cite{Celery}.\\ Number of connections is limited by 10 connections~\cite{Celery}.\end{tabular} &
  \begin{tabular}[c]{@{}l@{}}Enables operations to manage and maintain distributed task queues\\,such as starting, stopping, and restarting worker processes~\cite{Celery}.\\ Functions as a task queue~\cite{Celery}. \\Focuses on real-time processing~\cite{Celery}.\\ Supports task scheduling~\cite{Celery}.\\ Supports multi-message brokers~\cite{Celery}. \\ Integrates with multi-web frameworks~\cite{Celery}.\\ Supports automatic retry in the event of connection loss or \\failure~\cite{Celery}.\end{tabular} \\ \hline
\rowcolor[HTML]{F4EFEF} 
\textbf{JBoss Messaging~\cite{JBossMessaging}} &
  \cmark &
  \cmark &
  \cmark &
  \cmark &
  Push &
  \cmark &
  \cmark &
  \cmark &
  \cmark &
  \cmark &
  \cmark &
  \begin{tabular}[c]{@{}l@{}}Delay may occur with large messages (up to 100KB) due\\ to split message into multiple packages\cite{JBossCons}. \end{tabular} &
  \begin{tabular}[c]{@{}l@{}}Supports AMQP, MQTT, STOMP message protocols~\cite{JBossCons}.\\ Supports transactions~\cite{JBossCons}.\\ Provides management processes related to deployments, configuration, \\and access control~\cite{JBossCons}.\\ Easy integration with other JBoss and Java EE components~\cite{JBossCons}.\end{tabular} \\ \hline
\rowcolor[HTML]{FFFFFF} 
\textbf{OpenMQ~\cite{OpenMQ}} &
  \cmark &
  \cmark &
  \cmark &
  \cmark &
  Both &
  \cmark &
  \cmark &
  \cmark &
  \cmark &
  \cmark &
  \cmark &
  \begin{tabular}[c]{@{}l@{}}Setup is complex~\cite{OpenMQ}.\\ High Latency~\cite{donta2022survey}.\end{tabular} &
  Loosely-coupled architecture~\cite{OpenMQ}. \\ \hline
\rowcolor[HTML]{F4EFEF} 
\textbf{Beanstalk~\cite{Beanstalk}} &
  \cmark &
  \cmark &
  \xmark &
  \cmark &
  Pull &
  \cmark &
  \cmark &
  \xmark &
  \cmark &
  \xmark &
  \xmark &
  \begin{tabular}[c]{@{}l@{}}Lacks authentication~\cite{Beanstalk}.\\ Message size is limited to 64 KB~\cite{Beanstalkprotocol2}.\end{tabular} &
  \begin{tabular}[c]{@{}l@{}}Unprocessed messages are automatically returned to the queue~\cite{Beanstalkprotocol2}.\\ Supports Ruby, Rails, Java, JavaScript, Haskell, and PHP~\cite{Beanstalkprotocol}.\end{tabular} \\ \hline
\rowcolor[HTML]{FFFFFF} 
\textbf{Gearman~\cite{Gearman}} &
  \cmark &
  \cmark &
  \xmark &
  \cmark &
  Both &
  \xmark &
  \cmark &
  \xmark &
  \cmark &
  \cmark &
  \cmark &
  \begin{tabular}[c]{@{}l@{}}Does not have authentication and SSL support~\cite{Gearman}.\\ Manual configuration~\cite{Gearman}.\\ Monitoring tools are limited~\cite{Gearman}.\end{tabular} &
  \begin{tabular}[c]{@{}l@{}}Used by LiveJournal, Yahoo!, and Digg~\cite{Gearman}.\\ Multi-languages support~\cite{Gearman}. \\ No single point of failure~\cite{Gearman}. \\ No limits on message size~\cite{Gearman}. \\ Supports load balancing~\cite{Gearman}. \end{tabular} \\ \hline
\rowcolor[HTML]{F4EFEF} 
\textbf{Enduro/X~\cite{Enduro/X}} &
  \cmark &
  \cmark &
  \cmark &
  \cmark &
  Both &
  \cmark &
  \cmark &
  \xmark &
  \cmark &
  \cmark &
  \cmark &
  \begin{tabular}[c]{@{}l@{}}Message size is limited to max 10 MB~\cite{Mavimax2022Enduro/X1}.\\Buffer size is limited to max 64KB~\cite{Mavimax2022Enduro/X1}.\\Cluster nodes number is limited to max 32 nodes~\cite{Mavimax2022Enduro/X1}.\\ Resource managers numbers with single transaction \\are limited to max 32~\cite{Mavimax2022Enduro/X1}.\\ Versions compatibility depends on the date of \\release~\cite{Enduro/X}.\\ Limitations on availability of the operations that can be\\ executed within the callback~\cite{Mavimax2022Enduro/X1}. \end{tabular} &
  \begin{tabular}[c]{@{}l@{}}Distributed transaction processing~\cite{Mavimax2022Enduro/X1}.\\Works on multi-platforms~\cite{Enduro/X}. \end{tabular} \\ \hline
\rowcolor[HTML]{FFFFFF} 
\textbf{WSO2~\cite{WSO2}} &
  \cmark &
  \cmark &
  \cmark &
  \cmark &
  Both &
  \cmark &
  \cmark &
  \cmark &
  \cmark &
  \cmark &
  \cmark &
  \begin{tabular}[c]{@{}l@{}}Heap memory size allocation is limited to max \\4GB~\cite{WSO2memory}.\end{tabular} &
  \begin{tabular}[c]{@{}l@{}}Supports widely used  protocols such as HTTP/S, JMS, VFS, UDP,\\ TCP, MQTT, MSMQ, and MailTo~\cite{WSO2transport}.\\ Supports message filtering~\cite{WSO2filter}.\\ Integrates easily with other WSO2 products and third-party\\ systems~\cite{WSO2}.\end{tabular} \\ \hline
  \rowcolor[HTML]{F4EFEF} 
\textbf{HiveMQ~\cite{HiveMQ}} &
  \cmark &
  \cmark &
  \cmark &
  \cmark &
  Push &
  \cmark &
  \cmark &
  \cmark &
  \cmark &
  \cmark &
  \cmark &
  \begin{tabular}[c]{@{}l@{}}Number of characters broker accepts in an Client ID is \\limited between 1 and 65535~\cite{HiveMQ2}.\\ Number of characters broker accepts in a topic string is \\limited between 1 and 65535~\cite{HiveMQ2}.\\ Resource intensive for maintenance~\cite{HiveMQ2}.\end{tabular} &
  \begin{tabular}[c]{@{}l@{}}Is a client-based MQTT broker for M2M communication~\cite{HiveMQ}.\\ Suitable for mission-critical applications~\cite{HiveMQ}.\\ Supports real-time monitoring of device data \&  integration with \\existing systems~\cite{HiveMQ}.\end{tabular} \\ \hline
  \rowcolor[HTML]{FFFFFF} 
\textbf{Redis~\cite{Redis}} &
  \cmark &
  \cmark &
  \cmark &
  \cmark &
  Push &
  \cmark &
  \cmark &
  \cmark &
  \cmark &
  \cmark &
  \cmark &
  \begin{tabular}[c]{@{}l@{}}Uses a memory dump which leads to slow \\performance~\cite{Redis}.\\ Has only basic security options~\cite{Redis2}.\end{tabular} &
  \begin{tabular}[c]{@{}l@{}}Supports multiple data types~\cite{Redis}.\\ In-memory data storage~\cite{Redis}.\end{tabular} \\ \hline
\rowcolor[HTML]{F4EFEF} 
  \textbf{EMQX~\cite{EMQX}} &
  \cmark &
  \cmark &
  \cmark &
  \cmark &
  Both &
  \cmark &
  \cmark &
  \cmark &
  \cmark &
  \cmark &
  \cmark &
  \begin{tabular}[c]{@{}l@{}}Setup, configuration and management are complex~\cite{EMQX4}.\end{tabular} &
  \begin{tabular}[c]{@{}l@{}}Supports MQTT bridging~\cite{EMQX}.\\ Supports data integration~\cite{EMQX}.\end{tabular} \\ \hline
  \rowcolor[HTML]{FFFFFF} 
\textbf{Apache Pulsar~\cite{Pulsar}} &
  \cmark &
  \cmark &
  \cmark &
  \cmark &
  Push &
  \cmark &
  \cmark &
  \cmark &
  \cmark &
  \cmark &
  \cmark &
  \begin{tabular}[c]{@{}l@{}}Complex configuration and deployment~\cite{Pulsar}. \\ Complex architecture, based on four components (Pulsar\\ servers, Apache BookKeeper, Apache ZooKeeper, and the \\RocksDB database) that need to be configured and \\managed~\cite{Pulsar}.\end{tabular} &
  \begin{tabular}[c]{@{}l@{}}Supports event streaming~\cite{Pulsar}.\\ An index-based storage system~\cite{Pulsar}.\\ Low latency~\cite{Pulsar}. \\ Supports messaging, streaming, and queuing~\cite{Pulsar}.\end{tabular} \\ \hline
  \rowcolor[HTML]{F4EFEF} 
\end{tabular}%
}
\end{table*}

\begin{table*}[]
\centering
\caption{A summary of open source and non-priority-supporting message brokers.}
\label{tab:MBs2}
\resizebox{!}{.39\paperheight}{%
\begin{tabular}{|c|c|c|c|c|c|c|c|c|c|c|c|l|l|}
\hline
\rowcolor[HTML]{F4EFEF} 
\rotatebox{90}{\textbf{\begin{tabular}[c]{@{}c@{}}Message \\ Brokers and Queues\end{tabular}}} &
  \rotatebox{90}{\textbf{Clustering Support}} &
  \rotatebox{90}{\textbf{Monitoring Support}} &
  \rotatebox{90}{\textbf{Pub/Sub Support}} &
  \rotatebox{90}{\textbf{Parallel Processing}} &
  \rotatebox{90}{\textbf{Pull \& Push Support}} &
  \rotatebox{90}{\textbf{Reliable Delivery}} &
  \rotatebox{90}{\textbf{Persistent}} &
  \rotatebox{90}{\textbf{Authentication}} &
  \rotatebox{90}{\textbf{Scalable}} &
  \rotatebox{90}{\textbf{Distributed}} &
  \rotatebox{90}{\textbf{Fault Tolerance}} &
  \multicolumn{1}{c|}{\cellcolor[HTML]{F4EFEF}\rotatebox{90}{\textbf{Shortcomings}}} &
  \multicolumn{1}{c|}{\cellcolor[HTML]{F4EFEF}\rotatebox{90}{\textbf{Features}}} \\ \hline
\rowcolor[HTML]{FFFFFF} 
\textbf{Apache Kafka~\cite{Kafka}} &
  \cmark &
  \cmark &
  \cmark &
  \cmark &
  Pull &
  \cmark &
  \cmark &
  \cmark &
  \cmark &
  \cmark &
  \cmark &
  \begin{tabular}[c]{@{}l@{}}Resource intensive~\cite{Kafka}.\\ Provides a data backlog~\cite{Kafka}. \\ Complex~\cite{Kafka}.\end{tabular} &
  \begin{tabular}[c]{@{}l@{}}Supports topic (log) compaction \& distributed event streaming~\cite{Kafka}.\\Supports data integration~\cite{Kafka}.\\ Language support~\cite{Kafka}.  \\ Supports multiple data formats~\cite{Kafka}.\\ Supports permanent storage \& the management of data flow and\\ consumer groups~\cite{Kafka}.\\ Supports replication and partitioning of data~\cite{Kafka}.\\ Supports deployment in different environments~\cite{Kafka}.\end{tabular} \\ \hline
 \rowcolor[HTML]{F4EFEF}
  \textbf{Apache RocketMQ~\cite{ApacheRocketMQ}} &
  \cmark &
  \cmark &
  \cmark &
  \cmark &
  Both &
  \cmark &
  \cmark &
  \cmark &
  \cmark &
  \cmark &
  \cmark &
  \begin{tabular}[c]{@{}l@{}}Message size is limited to max 4MB~\cite{ApacheRocketMQ}.\\Message sending retries is limited to max 3 times~\cite{ApacheRocketMQ}.\end{tabular} &
  \begin{tabular}[c]{@{}l@{}}Supports message broadcasting, tracking, filtering, and retrying~\cite{ApacheRocketMQ}.\\Low latency~\cite{ApacheRocketMQ}.\\ Maintains the order of messages~\cite{ApacheRocketMQ}.\\ Supports multi-protocols~\cite{ApacheRocketMQ}.\\Supports multiple programming languages~\cite{ApacheRocketMQ}.\end{tabular} \\ \hline
\rowcolor[HTML]{FFFFFF} 
\textbf{Eclipse Mosquitto~\cite{EclipseMosquitto}} &
  \xmark &
  \cmark &
  \cmark &
  \xmark &
  Both &
  \cmark &
  \cmark &
  \cmark &
  \cmark &
  \cmark &
  \cmark &
  \begin{tabular}[c]{@{}l@{}}Limited security~\cite{EclipseMosquitto}.\\ No built-in clustering~\cite{EclipseMosquitto6}.\\ Unsuitable for large-scale deployments~\cite{EclipseMosquitto6}. \\ Deployment is challenging in a cloud environment~\cite{EclipseMosquitto6}.\\ Message size is limited to max 256MB~\cite{EclipseMosquitto}. \end{tabular} &
  \begin{tabular}[c]{@{}l@{}}Low resource usage~\cite{EclipseMosquitto3}.\\ QoS support~\cite{EclipseMosquitto}.\\ Topic-based message filtering~\cite{EclipseMosquitto}.\\ Supports logging and debugging~\cite{EclipseMosquitto}.\\ Supports functioning as a bridge~\cite{EclipseMosquitto}.\\ Supports dynamic restart configuration~\cite{EclipseMosquitto}.\\ Suitable for low-power machines~\cite{EclipseMosquitto3}.\end{tabular} \\ \hline
  \rowcolor[HTML]{F4EFEF} 
\textbf{ZeroMQ~\cite{zeromq}} &
  \cmark &
  \cmark &
  \cmark &
  \cmark &
  Both &
  \cmark &
  \xmark &
  \cmark &
  \cmark &
  \cmark &
  \cmark &
  \begin{tabular}[c]{@{}l@{}}High load of local control modules~\cite{donta2022survey}. \\ Fails to manage relationships between all network \\components~\cite{donta2022survey}. \\Limited security~\cite{zeromq3}.\\ Scaling is challenging~\cite{zeromq2}. \\ Delivery is not guaranteed~\cite{zeromq2}.\end{tabular} &
  \begin{tabular}[c]{@{}l@{}}Brokerless messaging platform~\cite{zeromq}.\\ Multi-languages and platforms support~\cite{zeromq}.\\ Carries messages across IPC, TCP, TPIC, and multicast~\cite{zeromq}.\\ Low latency~\cite{zeromq}.\end{tabular} \\ \hline
\rowcolor[HTML]{FFFFFF} 
\textbf{Apache NiFi~\cite{ApacheNiFi}} &
  \cmark &
  \cmark &
  \cmark &
  \cmark &
  Both &
  \cmark &
  \cmark &
  \cmark &
  \cmark &
  \cmark &
  \cmark &
  \begin{tabular}[c]{@{}l@{}}Data extraction is difficult when a node is \\separated from a cluster~\cite{ApacheNiFi}.\\ Under certain conditions, data is automatically \\deleted~\cite{ApacheNiFi}. \\ Complex configuration~\cite{ApacheNiFi}.\end{tabular} &
  \begin{tabular}[c]{@{}l@{}}Provides a data flow framework~\cite{ApacheNiFi}.\\ Provides data compression using a user-specified algorithm to reduce\\ data size~\cite{ApacheNiFi}.\\ Prevents data loss by controlling data flow and stopping the \\production of more data than a queue can handle~\cite{ApacheNiFi}.\\ Supports buffering of all queued data~\cite{ApacheNiFi}.\\ Integrates and processes multiple data sources~\cite{ApacheNiFi}. \end{tabular} \\ \hline
\rowcolor[HTML]{F4EFEF} 
\textbf{Ably Realtime~\cite{AblyRealtime}} &
  \cmark &
  \cmark &
  \cmark &
  \cmark &
  Both &
  \cmark &
  \cmark &
  \cmark &
  \cmark &
  \cmark &
  \cmark &
  \begin{tabular}[c]{@{}l@{}}Caps the number of channels per connection~\cite{AblyRealtime}. \\ Peak connections number limited between 200 and \\240~\cite{AblyRealtime}.\\ Message size limited to 16KB~\cite{AblyRealtime}. \\ Number of queues limited to 5~\cite{AblyRealtime}.\\ Queue length limited to 10,000~\cite{AblyRealtime}.\end{tabular} &
  \begin{tabular}[c]{@{}l@{}}Addresses challenging real-time requirements~\cite{AblyRealtime}.\\ Supports streaming data~\cite{AblyRealtime}. \\ Supports multiple protocols~\cite{AblyRealtime}.\end{tabular} \\ \hline
  \rowcolor[HTML]{FFFFFF} 
\textbf{Apache SamZa~\cite{SamZa}} &
  \cmark &
  \cmark &
  \cmark &
  \cmark &
  Pull &
  \cmark &
  \cmark &
  \xmark &
  \cmark &
  \cmark &
  \cmark &
  \begin{tabular}[c]{@{}l@{}}Only supports JVM languages~\cite{SamZa2}.\\ Configuration is complex~\cite{SamZa}.\\ Resources intensive with large data volumes~\cite{kleppmann2019apache}.\end{tabular} &
  \begin{tabular}[c]{@{}l@{}}Stream processing framework~\cite{SamZa}.\\ Supports message storage, routing, and processing management~\cite{SamZa}.\\ Supports at-least once data processing~\cite{SamZa}.\\ Real-time data processing with low latency~\cite{SamZa}. \\ Easy to integrate~\cite{SamZa}.\end{tabular} \\ \hline
\rowcolor[HTML]{F4EFEF} 
\textbf{VerneMQ~\cite{VerneMQ}} &
  \cmark &
  \cmark &
  \cmark &
  \cmark &
  Both &
  \cmark &
  \cmark &
  \cmark &
  \cmark &
  \cmark &
  \cmark &
  \begin{tabular}[c]{@{}l@{}}Lack of security~\cite{donta2022survey}.\\ Clustering architecture is unproofed~\cite{EMQX5}.\\ Limited enterprise features~\cite{EMQX5}.\\ Not under active development~\cite{EMQX5}.\\ Limited support for MQTT integration~\cite{EMQX5}.\\ Lacks management and monitoring features~\cite{EMQX5}.\\ No cloud-based service~\cite{VerneMQ}.\end{tabular} &
  \begin{tabular}[c]{@{}l@{}}Master-less clustered messaging protocol~\cite{VerneMQ}.\\ Supports flow control~\cite{donta2022survey}.\\ Low latency~\cite{VerneMQ}.\end{tabular} \\ \hline
  \rowcolor[HTML]{FFFFFF} 
\textbf{NServiceBus~\cite{NServiceBus}} &
  \cmark &
  \cmark &
  \cmark &
  \cmark &
  - &
  \cmark &
  \cmark &
  \xmark &
  \cmark &
  \cmark &
  \cmark &
  \begin{tabular}[c]{@{}l@{}}Scalability is limited due to use centralized \\resource~\cite{NServiceBus1}. \\ Monitoring tools are limited~\cite{NServiceBus4}. \\ Debugging is complex with huge stream of \\messages~\cite{NServiceBus1}.\end{tabular} &
  \begin{tabular}[c]{@{}l@{}}Ensures message processing~\cite{NServiceBus}.\\ Supports transactions and recovery is built-in~\cite{NServiceBus1}.\\ Messages can be retried at regular intervals~\cite{NServiceBus1}.\end{tabular} \\ \hline
\rowcolor[HTML]{F4EFEF} 
\textbf{Kestrel~\cite{kestrel}} &
  \xmark &
  \cmark &
  \xmark &
  \cmark &
  Pull &
  \cmark &
  \cmark &
  \xmark &
  \xmark &
  \xmark &
  \xmark &
  \begin{tabular}[c]{@{}l@{}}Low support of security~\cite{kestrel}.\\ Low clustering capabilities~\cite{kestrel}.\\ Memory size is limited to max 128MB~\cite{kestrel}.\\ Number of items in the queue is limited to 500~\cite{kestrel}. \\ Data size of each item in the queue is limited to max \\32bytes~\cite{kestrel2}.\end{tabular} &
  \begin{tabular}[c]{@{}l@{}}Written in Scala~\cite{kestrel}.\\ Each server handles ordered MQs, with no cross \\communication, resulting in a cluster of k-ordered queues~\cite{kestrel}.\end{tabular} \\ \hline
\rowcolor[HTML]{FFFFFF} 
\textbf{NSQ~\cite{NSQ}} &
  \cmark &
  \cmark &
  \cmark &
  \cmark &
  Push &
  \cmark &
  \cmark &
  \xmark &
  \cmark &
  \cmark &
  \cmark &
  \begin{tabular}[c]{@{}l@{}}Data loss with server crash~\cite{NSQ}.\\ Messages are unordered~\cite{NSQ}. \\ Limited persistence~\cite{NSQ}.\\ No message recovery~\cite{NSQ}.\\ Lacks replication~\cite{NSQ}. \\ Messages are delivered at least once, which may \\duplicate messages~\cite{NSQ}.\end{tabular} &
  \begin{tabular}[c]{@{}l@{}}Load-balanced message delivery~\cite{NSQ}. \\ Efficient handling of high-volume\& real-time data streams~\cite{NSQ}.\end{tabular} \\ \hline
\rowcolor[HTML]{F4EFEF} 
\textbf{NATS~\cite{NATS}} &
  \cmark &
  \cmark &
  \cmark &
  \cmark &
  Both &
  \cmark &
  \cmark &
  \cmark &
  \cmark &
  \cmark &
  \cmark &
  Message size is limited to max 64MB~\cite{NATS1}. &
  \begin{tabular}[c]{@{}l@{}}Suitable for real-time communication~\cite{NATS1}.\\ Easy to use~\cite{NATS1}.\\ Minimal resource consumption~\cite{NATS1}. \\ Offers persistence with "at-least-once" and "exactly-once"~\cite{NATS1}.\end{tabular} \\ \hline
\rowcolor[HTML]{FFFFFF} 
\textbf{KubeMQ~\cite{KubeMQ}} &
  \cmark &
  \cmark &
  \cmark &
  \cmark &
  Both &
  \cmark &
  \cmark &
  \cmark &
  \cmark &
  \cmark &
  \cmark &
  \begin{tabular}[c]{@{}l@{}}Unsuitable to all use cases due to it's designed for \\dynamic microservice environments~\cite{ayaz2022data}.\end{tabular} &
  \begin{tabular}[c]{@{}l@{}}Builds a hybrid infrastructure across clouds, on-prem, and at the\\ edge to allow microservices from multi-environments to\\ communicate~\cite{KubeMQ}.\\ Support for pub/sub, microservices, multistage pipeline, \\and tasks queue use cases~\cite{KubeMQ}. \\ Runs in Kubernetes and connects natively to the K8S\\ cloud-native ecosystem~\cite{KubeMQ}.\\ Simple deployment in Kubernetes~\cite{KubeMQ}.\\ Easy to use~\cite{KubeMQ}.\\Low latency~\cite{KubeMQ}. \end{tabular} \\ \hline
\end{tabular}%
}
\end{table*}
\begin{table*}[]
\centering
\caption{Summary of proprietary and priority-supporting message brokers.}
\label{tab:MBs3}
\resizebox{!}{.3\paperheight}{%
\begin{tabular}{|c|c|c|c|c|c|c|c|c|c|c|c|l|l|}
\hline
\rowcolor[HTML]{F4EFEF} 
\rotatebox{90}{\textbf{\begin{tabular}[c]{@{}c@{}}Message \\ Brokers and Queues\end{tabular}}} &
  \rotatebox{90}{\textbf{Clustering Support}} &
  \rotatebox{90}{\textbf{Monitoring Support}} &
  \rotatebox{90}{\textbf{Pub/Sub Support}} &
  \rotatebox{90}{\textbf{Parallel Processing}} &
  \rotatebox{90}{\textbf{Pull \& Push Support}} &
  \rotatebox{90}{\textbf{Reliable Delivery}} &
  \rotatebox{90}{\textbf{Persistent}} &
  \rotatebox{90}{\textbf{Authentication}} &
  \rotatebox{90}{\textbf{Scalable}} &
  \rotatebox{90}{\textbf{Distributed}} &
  \rotatebox{90}{\textbf{Fault Tolerance}} &
  \multicolumn{1}{c|}{\cellcolor[HTML]{F4EFEF}\rotatebox{90}{\textbf{Shortcomings}}} &
  \multicolumn{1}{c|}{\cellcolor[HTML]{F4EFEF}\rotatebox{90}{\textbf{Features}}} \\ \hline
\rowcolor[HTML]{FFFFFF} 
\textbf{IBM MQ~\cite{ibmmq}} &
  \cmark &
  \cmark &
  \cmark &
  \cmark &
  Push &
  \cmark &
  \cmark &
  \cmark &
  \cmark &
  \cmark &
  \cmark &
  \begin{tabular}[c]{@{}l@{}}High costs~\cite{ibmmq1}.\\Problems with message prioritization due to unordered way \\of allocating messages~\cite{IronMQibm}. \\Does not always integrate with the newest forms of \\messaging~\cite{IronMQibm}.\\Messages are unordered~\cite{IronMQibm}.\end{tabular} &
  \begin{tabular}[c]{@{}l@{}}QoS support~\cite{ibmmq9}.\\  Provides robust monitoring \& tracing of all\\ messages~\cite{ibmmq9}.\\ Controls undelivered messages~\cite{ibmmq9}.\\ Multi-APIs support~\cite{ibmmq9}.\\ Allows applications to be decoupled~\cite{ibmmq9}.\\ Easy to deploy on various platforms~\cite{ibmmq9}.\end{tabular} \\ \hline
\rowcolor[HTML]{F4EFEF} 
\textbf{Amazon SQS~\cite{SQS}} &
  \xmark &
  \cmark &
  \cmark &
  \cmark &
  Pull &
  \cmark &
  \cmark &
  \cmark &
  \cmark &
  \cmark &
  \cmark &
  \begin{tabular}[c]{@{}l@{}}High scale-up cost~\cite{SQS}.\\ Message size is limited between 1KB and 256 KB~\cite{SQS3}.\\ Message ordering is not guaranteed~\cite{SQS}.\\ Message retention before deletion is limited between 1 minute \\and 14 days~\cite{SQS}.\\\end{tabular} &
  \begin{tabular}[c]{@{}l@{}}Cloud-based web service~\cite{SQS}. \\ Supports decoupling microservices, distributed \\systems, and serverless applications~\cite{SQS}. \\ Transmits, stores, and receives messages across \\software components using SQS at any volume~\cite{SQS}.\\ Messages are delivered at least once~\cite{SQS}.\end{tabular} \\ \hline
\rowcolor[HTML]{FFFFFF} 
\textbf{Microsoft MQ (MSMQ)~\cite{MSMQ}} &
  \cmark &
  \cmark &
  \cmark &
  \cmark &
  Push &
  \cmark &
  \cmark &
  \cmark &
  \cmark &
  \cmark &
  \cmark &
  \begin{tabular}[c]{@{}l@{}}May experience resource failure~\cite{MSMQ}.\\ Limitation on message size~\cite{MSMQ}.\\ Drops all MSMQ messages if the appropriate server is not \\deployed~\cite{MSMQ7}. \\ Open queue failure error prevents data transfer~\cite{MSMQ7}.\end{tabular} &
  \begin{tabular}[c]{@{}l@{}}Multi-protocols support~\cite{MSMQ}.\\ Tracks and deletes expired messages~\cite{MSMQ}.\\ Manages distributed brokers~\cite{MSMQ}.\\ Supports remote access~\cite{MSMQ}. \\ Effective routing~\cite{MSMQ}.\\ Provides guaranteed message delivery~\cite{MSMQ}.\\ Provides a store and forward mechanism~\cite{MSMQ}. \\ Supports transactions~\cite{MSMQ}.\end{tabular} \\ \hline
\rowcolor[HTML]{F4EFEF} 
\textbf{\begin{tabular}[c]{@{}c@{}}Oracle GlassFish Server \\ Message Queue~\cite{Oracle}\end{tabular}} &
  \cmark &
  \cmark &
  \cmark &
  \cmark &
  Both &
  \cmark &
  \cmark &
  \cmark &
  \cmark &
  \cmark &
  \cmark &
  \begin{tabular}[c]{@{}l@{}}Resource intensive as the number of messages increases~\cite{Oracle}.\\ High latency as connections number to the broker \\increases~\cite{Oracle}. \\ Size of message is limited to max 70MB~\cite{Oracle}.\end{tabular} &
  \begin{tabular}[c]{@{}l@{}}Supports transactions~\cite{Oracle}.\\ Supports JMS 1.1, STOMP, and HTTP \\protocols\cite{Oracle}.\\ Well-known standards-based messaging \\support~\cite{Oracle}.\end{tabular} \\ \hline
\rowcolor[HTML]{FFFFFF} 
\textbf{\begin{tabular}[c]{@{}c@{}}TIBCO \\ Enterprise \\ Message Service~\cite{TIBCOEnterpriseMessageService}\end{tabular}} &
  \cmark &
  \cmark &
  \cmark &
  \cmark &
  Both &
  \cmark &
  \cmark &
  \cmark &
  \cmark &
  \cmark &
  \cmark &
  \begin{tabular}[c]{@{}l@{}}Excludes fault tolerance of the server~\cite{TIBCO1}.\\ Recourse intensive as message size increased up to 512MB~\cite{TIBCO3}.\end{tabular} &
  \begin{tabular}[c]{@{}l@{}}Supports load balancing~\cite{TIBCO1}.\\ Manages the real-time flow of information~\cite{TIBCOEnterpriseMessageService}. \\ Supports multiple message protocols and \\technologies~\cite{TIBCO1}.\\ Easy integration with TIBCO eFTL™ software\\ expands broker to web and mobile\\ applications~\cite{TIBCOEnterpriseMessageService}. \\ Loosely coupled design~\cite{TIBCOEnterpriseMessageService}.\\ Supports integration for heterogeneous \\platforms~\cite{TIBCOEnterpriseMessageService}.\end{tabular} \\ \hline
\rowcolor[HTML]{F4EFEF} 
\textbf{TIBCO Rendezvous~\cite{TIBCORendezvous}} &
  \xmark &
  \cmark &
  \cmark &
  \cmark &
  Both &
  \cmark &
  \cmark &
  \cmark &
  \cmark &
  \cmark &
  \cmark &
  \begin{tabular}[c]{@{}l@{}}Expensive~\cite{TIBCORendezvous3}.\\ Queue size limited to max 500~\cite{TIBCORendezvous2}.\end{tabular} &
  \begin{tabular}[c]{@{}l@{}}Supports C, C++, Java , \& .NET programming \\language~\cite{TIBCORendezvous1}.\\ Easy to use \& setup~\cite{TIBCORendezvous1}.\\ Has a distributed architecture to eliminate\\ failure~\cite{TIBCORendezvous1}.\end{tabular} \\ \hline
\rowcolor[HTML]{FFFFFF} 
\textbf{Anypoint MQ~\cite{AnypointMQ}} &
  \cmark &
  \cmark &
  \cmark &
  \cmark &
  Pull &
  \cmark &
  \cmark &
  \cmark &
  \cmark &
  \cmark &
  \cmark &
  \begin{tabular}[c]{@{}l@{}}Expensive~\cite{AnypointMQ3}.\\ Payload size is limited to max 10 MB~\cite{AnypointMQ}.\\ Converts the payload format, leading to an increase in\\payload size.~\cite{AnypointMQ}.\end{tabular} &
  \begin{tabular}[c]{@{}l@{}}Supports data integration~\cite{AnypointMQ3}.\\ Stores messages in a queue~\cite{AnypointMQ}. \\ Provides intelligent message routing~\cite{AnypointMQ}.\end{tabular} \\ \hline
\rowcolor[HTML]{F4EFEF} 
\textbf{Azure Service Bus~\cite{AzureServiceBus}} &
  \cmark &
  \cmark &
  \cmark &
  \cmark &
  Both &
  \cmark &
  \cmark &
  \cmark &
  \cmark &
  \cmark &
  \cmark &
  \begin{tabular}[c]{@{}l@{}}Is a cloud-only service~\cite{AzureServiceBus}.\\Message size is limited between 256 KB and 100 MB~\cite{AzureServiceBus}.\\ Number of queues is to max 10,000~\cite{Azure}. \\ Number of subscriptions per topic is limited to max 2,000~\cite{AzureServiceBus}.\end{tabular} &
  \begin{tabular}[c]{@{}l@{}}Provides duplicate detection, duplicate\\ messages will not be stored in the queue~\cite{AzureServiceBus}.\\ Guarantees ordering~\cite{Azure}. \\ Offers scheduling~\cite{AzureServiceBus5,AzureServiceBus4}.\\ Integrates well with other Azure products~\cite{AzureServiceBus4}.\\ Supports multi-protocols~\cite{AzureServiceBus}.\\ Provides delivery guarantee (at-least-once, \\at-most-once)~\cite{Azure}.\end{tabular} \\ \hline
\rowcolor[HTML]{FFFFFF} 
\textbf{SAP NW PI~\cite{SAP}} &
  \cmark &
  \cmark &
  \cmark &
  \cmark &
  Both &
  \cmark &
  \cmark &
  \cmark &
  \cmark &
  \cmark &
  \cmark &
  \begin{tabular}[c]{@{}l@{}} Message size is limited to max 350 MB~\cite{SAP3}.\\ Performance is directly affected by the message size~\cite{SAP3}. \end{tabular} &
  \begin{tabular}[c]{@{}l@{}}Supports various integration patterns~\cite{SAP}.\\ Supports message transformation~\cite{SAP}.\end{tabular} \\ \hline
\rowcolor[HTML]{F4EFEF} 
\textbf{Solace PubSub~\cite{Solace}} &
  \cmark &
  \cmark &
  \cmark &
  \cmark &
  Both &
  \cmark &
  \cmark &
  \cmark &
  \cmark &
  \cmark &
  \cmark &
  \begin{tabular}[c]{@{}l@{}}Message size is limited to 64 MB~\cite{Solace3}.\\ Broker connections is limited to max 1000, some functions \\are not available with the default 100 connections~\cite{Solace3}.\\ Guaranteed messaging is not supported over connections~\cite{Solace3}.\end{tabular} &
  \begin{tabular}[c]{@{}l@{}}Provides dynamic message routing~\cite{Solace1}.\\ High availability \& high-performance~\cite{Solace1}.\\ Provides distributed tracing~\cite{Solace}.\\ Event-driven architecture~\cite{Solace1}.\\ Supports multi-protocols~\cite{Solace1}.\end{tabular} \\ \hline
\end{tabular}%
}
\end{table*}

\begin{table*}[]
\centering
\caption{Summary of proprietary and non-priority-supporting message brokers.}
\label{tab:MBs4}
\resizebox{!}{.3\paperheight}{%
\begin{tabular}{|c|c|c|c|c|c|c|c|c|c|c|c|l|l|}
\hline
\rowcolor[HTML]{F4EFEF} 
\rotatebox{90}{\textbf{\begin{tabular}[c]{@{}c@{}}Message \\ Brokers and Queues\end{tabular}}} &
  \rotatebox{90}{\textbf{Clustering Support}} &
  \rotatebox{90}{\textbf{Monitoring Support}} &
  \rotatebox{90}{\textbf{Pub/Sub Support}} &
  \rotatebox{90}{\textbf{Parallel Processing}} &
  \rotatebox{90}{\textbf{Pull \& Push Support}} &
  \rotatebox{90}{\textbf{Reliable Delivery}} &
  \rotatebox{90}{\textbf{Persistent}} &
  \rotatebox{90}{\textbf{Authentication}} &
  \rotatebox{90}{\textbf{Scalable}} &
  \rotatebox{90}{\textbf{Distributed}} &
  \rotatebox{90}{\textbf{Fault Tolerance}} &
  \multicolumn{1}{c|}{\cellcolor[HTML]{F4EFEF}\rotatebox{90}{\textbf{Shortcomings}}} &
  \multicolumn{1}{c|}{\cellcolor[HTML]{F4EFEF}\rotatebox{90}{\textbf{Features}}} \\ \hline
\rowcolor[HTML]{FFFFFF} 
\textbf{\begin{tabular}[c]{@{}c@{}}Google Cloud\\  Pub/Sub~\cite{GoogleCloudPub/Sub}\end{tabular}} &
  \xmark &
  \cmark &
  \cmark &
  \cmark &
  Both &
  \cmark &
  \cmark &
  \cmark &
  \cmark &
  \cmark &
  \cmark &
  \begin{tabular}[c]{@{}l@{}}Unsuitable for large-scale deployments\\ due to limitations on resources~\cite{GoogleCloudPub/Sub}.\\ Message size is limited to  max 10MB~\cite{GoogleCloudPub/Sub}.\end{tabular} &
  \begin{tabular}[c]{@{}l@{}}Provides real-time stream analytic~\cite{GoogleCloudPub/Sub}.\\    Handles the underlying infrastructure, \\including provisioning servers, monitoring,\\ scaling, backups, and security updates~\cite{GoogleCloudPub/Sub1}. \\ Provides service maintenance feature that suitable \\for Google's most fundamental products to serve \\all customers effectively~\cite{GoogleCloudPub/Sub}. \\ Provides system maintenance feature to detect \\any issues with releases by continuously-running \\tests it is before used by customers and by \\monitoring~\cite{GoogleCloudPub/Sub}. \\ Integrates with other Google Cloud services~\cite{GoogleCloudPub/Sub}. \\ Supports automatic retries and message \\ordering~\cite{GoogleCloudPub/Sub}.\end{tabular} \\ \hline
\rowcolor[HTML]{F4EFEF} 
\textbf{Azure Storage Queue~\cite{AzureStorageQueue}} &
  \xmark &
  \cmark &
  \cmark &
  \cmark &
  Pull &
  \cmark &
  \cmark &
  \cmark &
  \cmark &
  \cmark &
  \cmark &
  \begin{tabular}[c]{@{}l@{}}Orders messages randomly~\cite{Azure}.\\Message size is limited to max 64 KB~\cite{AzureStorageQueue}.\end{tabular} &
  \begin{tabular}[c]{@{}l@{}}Maximum number of queues is unlimited~\cite{Azure}.\\ Activity monitoring support~\cite{AzureStorageQueue}.\\ Supports storing large numbers of messages~\cite{AzureStorageQueue}.\\Messages are delivered at-least-once~\cite{Azure}.\\ Does not provide duplicate detection~\cite{Azure}.\end{tabular} \\ \hline
\rowcolor[HTML]{FFFFFF} 
\textbf{Amazon MQ~\cite{AmazonMQ}} &
  \cmark &
  \cmark &
  \cmark &
  \cmark &
  Both &
  \cmark &
  \cmark &
  \cmark &
  \cmark &
  \cmark &
  \cmark &
  \begin{tabular}[c]{@{}l@{}}Number of broker connections \\is limited to 1,000, or 100 for micro \\brokers~\cite{AmazonMQ2}.\end{tabular} &
  \begin{tabular}[c]{@{}l@{}}Supports Standard Java Message Service (JMS) \\features~\cite{AmazonMQ2}.\\ Performs maintenance to the hardware, \\operating system,\& the engine software \\a message broker~\cite{AmazonMQ2}.\\ Integrates with other AWS services and \\applications~\cite{AmazonMQ}.\\ Supports distributed transactions~\cite{AmazonMQ2}.\\ Multi-protocols support~\cite{AmazonMQ2}.\end{tabular} \\ \hline
\rowcolor[HTML]{F4EFEF} 
\textbf{Intel MPI Library~\cite{IntelMPILibrary}} &
  \cmark &
  \cmark &
  \xmark &
  \cmark &
  Both &
  \cmark &
  \cmark &
  \cmark &
  \cmark &
  \cmark &
  \cmark &
  \begin{tabular}[c]{@{}l@{}}Expensive~\cite{IntelMPILibrary2}. \\ Compatibility issues with some \\systems~\cite{IntelMPILibrary,IntelMPILibrary1}.\\ Other processor architectures are not \\supported~\cite{IntelMPILibrary}.\end{tabular} &
  \begin{tabular}[c]{@{}l@{}}Uses OpenFabrics Interface (OFI) to handle all\\ communications~\cite{IntelMPILibrary}.\\ Establishes the connection only when needed, \\which reduces the memory footprint~\cite{IntelMPILibrary}. \\ Chooses the fastest transport available~\cite{IntelMPILibrary}. \\ Supports multi-cloud platforms~\cite{IntelMPILibrary}.\\ Supports multi-cluster interconnects~\cite{IntelMPILibrary}.\end{tabular} \\ \hline
\rowcolor[HTML]{FFFFFF} 
\textbf{Amazon Kinesis~\cite{Kinesis}} &
  \cmark &
  \cmark &
  \cmark &
  \cmark &
  Push &
  \cmark &
  \cmark &
  \cmark &
  \cmark &
  \cmark &
  \cmark &
  \begin{tabular}[c]{@{}l@{}}Permission issues~\cite{Kinesis5}.\\ Costly as data volume increases~\cite{Kinesis}.\\ Data payload size is limited to max\\ 1MB), data read rate to to 1MB/s, \\and number of consumers for each \\data stream to max 20~\cite{Kinesis6}. \\ Operations are rate-limited~\cite{Kinesis6}.\\ Limitation on number of data\\ streams~\cite{Kinesis6}. \\ Each record is added to a buffer with a \\deadline~\cite{Kinesis6}. \end{tabular} &
  \begin{tabular}[c]{@{}l@{}}Provides buffering and processing of real-time \\data streaming~\cite{Kinesis}.\\ Serverless streaming data service~\cite{Kinesis}.\\ Provides reliable data processing and delivery with\\ checkpointing and error-handling~\cite{Kinesis,Kinesis3}.\\ Offers monitoring and management~\cite{Kinesis}.\\ Offers various developer tools~\cite{Kinesis}.\end{tabular} \\ \hline
\rowcolor[HTML]{F4EFEF} 
\textbf{IronMQ~\cite{IronMQ}} &
  \cmark &
  \cmark &
  \cmark &
  \cmark &
  Both &
  \cmark &
  \cmark &
  \cmark &
  \cmark &
  \cmark &
  \cmark &
  \begin{tabular}[c]{@{}l@{}}Expensive~\cite{IronMQ2}.\\ Limited control~\cite{IronMQ1}.\end{tabular} &
  \begin{tabular}[c]{@{}l@{}}Meets the needs of both small businesses and large \\enterprises~\cite{IronMQ1}.\\ Supports multiple programming languages~\cite{IronMQ1}. \\ Uses REST API~\cite{IronMQ}. \\Easy to install~\cite{IronMQ1}. \\ Handles load buffering, synchronicity, and database \\offloading issues~\cite{IronMQ1}.\\ No limitation on the number of queues~\cite{IronMQ}.\end{tabular} \\ \hline

\end{tabular}%
}
\end{table*}
\subsection{Summary on Message Brokers}

Message brokers play a major role in streamlining communication between distributed systems by ensuring messages are properly routed. Among the features which they provide are guaranteed message delivery, queuing of messages, and pub/sub mechanisms. Additionally, they often provide mechanisms for message persistence so that they do not lose a single message in the event of a system failure. Their support for multiple messaging patterns, many different protocols, programming languages, and data styles, meets the needs of various types of communication. Moreover, many brokers come with an array of features. These features are critical elements, each contributing significantly to creating a robust message broker capable of managing communications in complex systems, ensuring efficiency. 

\textit{Clustering support}~\cite{ApacheActiveMQ} enables scaling the message service to accommodate more clients or connections, effectively handling large message volumes and numerous clients. \textit{Monitoring}~\cite{OpenMQmonitor} tools are crucial for tracking a message broker's performance and health, allowing for proactive management, early problem detection, and reliable operation. 

\textit{Pub/sub support}~\cite{ibmmq9} enables separation of publishers from subscribers, increasing system flexibility. \textit{Parallel processing support}~\cite{Kafka} allows the message broker to handle multiple messages simultaneously, improving throughput and efficiency. 

\textit{Pull and push support}~\cite{Kafka} allow flexible and timely message delivery. \textit{Reliable delivery support}~\cite{Oracle5} ensures messages are not lost during transit, typically through acknowledgments, retries, or temporary storage until successful delivery. 

\textit{Persistence support}~\cite{Oracle4} safeguards messages against loss during broker restarts or storage message failures. \textit{Authentication support}~\cite{Oracle4} is vital for permitting only authorized users and systems to publish or subscribe to messages, especially in systems dealing with sensitive data. 

\textit{Scalability}~\cite{RedHat3} refers to a message broker's ability to handle increasing loads from a large number of concurrently connected clients. Message brokers can operate in \textit{distributed environments}~\cite{ibmmq9}, allowing them to work on multiple devices or locations, optimizing workload distribution. \textit{Fault tolerance support}~\cite{ibmmq9} enables the message broker to recover from failures and continue operating smoothly.

With these capabilities, they are able to serve as the backbone for a wide variety of event-driven architectures and asynchronous communication patterns by serving as the basis for these architectures.

Despite their inherent benefits, message brokers have several limitations that must be addressed. As well as the difficulties these brokers face in deployment, maintenance, monitoring, integration, and management, they may cause considerable resource consumption. Furthermore, the limited ability of some brokers to handle large amounts of data make them unsuitable for handling real-time applications that are increasingly common in today's world.

\section{Message brokers and GenAI} \label{sec:ways}

\begin{figure*}[t]
\centering
\includegraphics[width=1\textwidth]{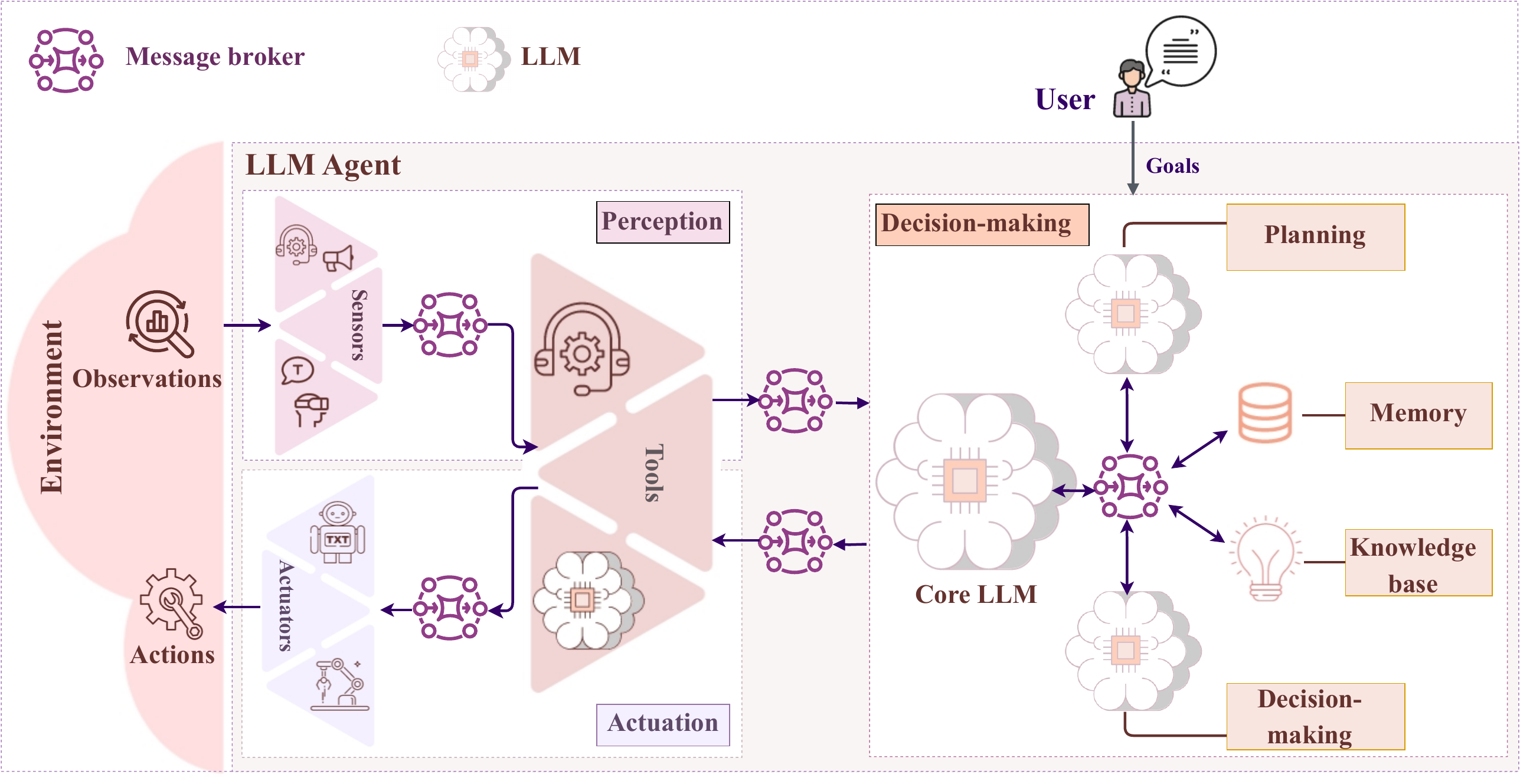}
\caption{The overall architecture of a GenAI agent, with possible integration points with message brokers.}\label{fig:Genai app}
\end{figure*}

The exploration of pub/sub communication patterns from the perspective of GenAI opens up a vision for the future, suggesting a range of benefits that could potentially enhance content delivery, personalization, and user engagement. Leveraging AI models such as GPT-3 \cite{brown2020language} and its successors holds the promise of delivering customized content in real-time, tailored to the unique preferences of individual subscribers. This could be achieved by automatically generating human-like text that aligns with each subscriber's interests~\cite{wang2023overview}. While the full realization of these benefits remains a subject for further research and development, the integration of advanced AI technologies with pub/sub systems offers a promising opportunities into the possibilities for more dynamic and personalized communication strategies. In this and following sections, we will attempt to shed light on this vision with practical examples and discuss how emerging enabling technologies can play a key role in this respect.




Language-based GenAI systems offers the capability to process subscriber queries and feedback efficiently using natural language understanding. These systems can power chatbots and virtual assistants, enabling users to communicate both interactively and intelligently with each other. Furthermore, their ability to summarize content, translate it into different languages, and moderate it significantly enhances the quality and accessibility of information~\cite{garcia2023we}. Consequently, such systems enable the design of proactive information delivery mechanisms, including automated reporting, anomaly detection, and predictive analytics~\cite{10198233}, thereby contributing to scalability, resource optimization, and load balancing \cite{wang2023overview}.

These advancements in language processing and interaction capabilities signify a critical opportunity in the evolution of computing architectures, especially as we address the rising processing and performance demands of GenAI applications. The complexity and sophistication of these applications necessitate a robust architectural framework that is not only capable of supporting the intricate dynamics of GenAI operations but also adaptable to the novel types of data generated by GenAI applications. This architecture should be specifically tailored to leverage the unique advantages that GenAI offers, such as enhanced content personalization and user engagement, while still aligning with traditional scenarios where message broker technologies are pivotal. For instance, the architecture can integrate LLMs and Large Multi-modal Models (LMM) to enhance the capabilities of actuators and sensors, enabling them to extract more semantic information from data and identify combinational patterns among them \cite{xu2024large}. This approach ensures that the system can dynamically adapt and respond to the evolving landscape of GenAI-driven communication, making it possible to abstract richer, more meaningful insights and foster synergistic interactions within the pub/sub ecosystem

In this regards, central to our discussion is the conceptualization of the \textit{GenAI agent model}, which exemplifies the architecture required to harness the full potential of GenAI applications \cref{fig:Genai app}. This model, segmented into \textit{environment} interaction, \textit{perception}, \textit{decision-making}, and \textit{ actuation components}, serves as the backbone for integrating GenAI capabilities with pub/sub systems. It encapsulates the essence of GenAI’s interaction with its surroundings, leveraging advanced computational engines like LLMs for processing and decision-making tasks.

The \textbf{perception component} perceives the environment through observations. It is equipped with sensing elements which may include physical sensors to gather diverse data from the surrounding environment, software-based tools, or both, along with a message broker designed to facilitate communication between these elements. These physical sensors can capture multi-modal observations, including visual, auditory, and textual data, as well as other modalities that can help the GenAI agent to understand its situation. The software-based tools such as LLMs, interface with abstract data streams. They read, analyze, and transform the gathered data into a comprehensible format for the agent's brain. These tools includes such as Multimodal-GPT~\cite{gong2023multimodal}, Flamingo~\cite{alayrac2022flamingo}, HuggingGPT~\cite{shen2023hugginggpt}, AudioGPT \cite{huang2023audiogpt}, GPT-4~\cite{openai2023gpt4}, Visual ChatGPT~\cite{wu2023visual}, etc. Using sensors with LLMs enhances their capability to understand and react to changes in the environment, leading to more effective decision-making in dynamic situations.

The agent's \textbf{decision-making component}, or brain, executes memorizing, thinking, analyzing and decision-making tasks, supporting long and short-term memory, knowledge, and planning~\cite{xi2023rise}. Short-term memory records recent tasks and actions, while long-term memory acts as an external database, enhancing the agent's ability to recall past conversations and pertinent details. Utilizing subgoal decomposition~\cite{wang2023survey} and a chain-of-thought approach~\cite{wei2022chain}, the agent breaks down large tasks into multiple manageable subgoals that are processed by a group of LLMs models. Through self-critics and reflection~\cite{yao2022react,shinn2023reflexion}, the agent can learn from its errors, enhancing its capabilities iteratively.

The agent uses actual physical actuators, LLMs-based tools, or both to execute tasks in the \textbf{actuation component}. These elements allow agents to interact with and respond to their environments. The LLMs-based tools include text generation through text-based tools such as ChatGPT~\cite{chatgpt2023}. Moreover, agents' workspaces have been expanded with embodied actions to support their integration and interaction with the physical world. LM-Nav~\cite{shah2023lm} analyzes input commands and the environment, aiming to identify the optimal walk based on a topological graph that is constructed internally. EmbodiedGPT~\cite{mu2023embodiedgpt} enables robots to comprehend and perform motion sequences in physical settings through multimodal visual understanding. Using these non-textual output tools extend the functionality of language models and agent scenarios. 

Crucially, GenAI agents are also able to generate novel tools. With frameworks like \textsc{CREATOR}~\cite{qian2023creator}, agents can generate executable programs or merge existing tools into more robust ones. Furthermore, with frameworks such as Self-Debugging \cite{chen2023teaching}, agents can iteratively improve the quality of the generated tools, autonomously learning from past experience, self-correcting and adapting, enhancing their tool-generation capabilities.

\begin{figure*}[h!]
\centering
\includegraphics[width=0.8\textwidth]{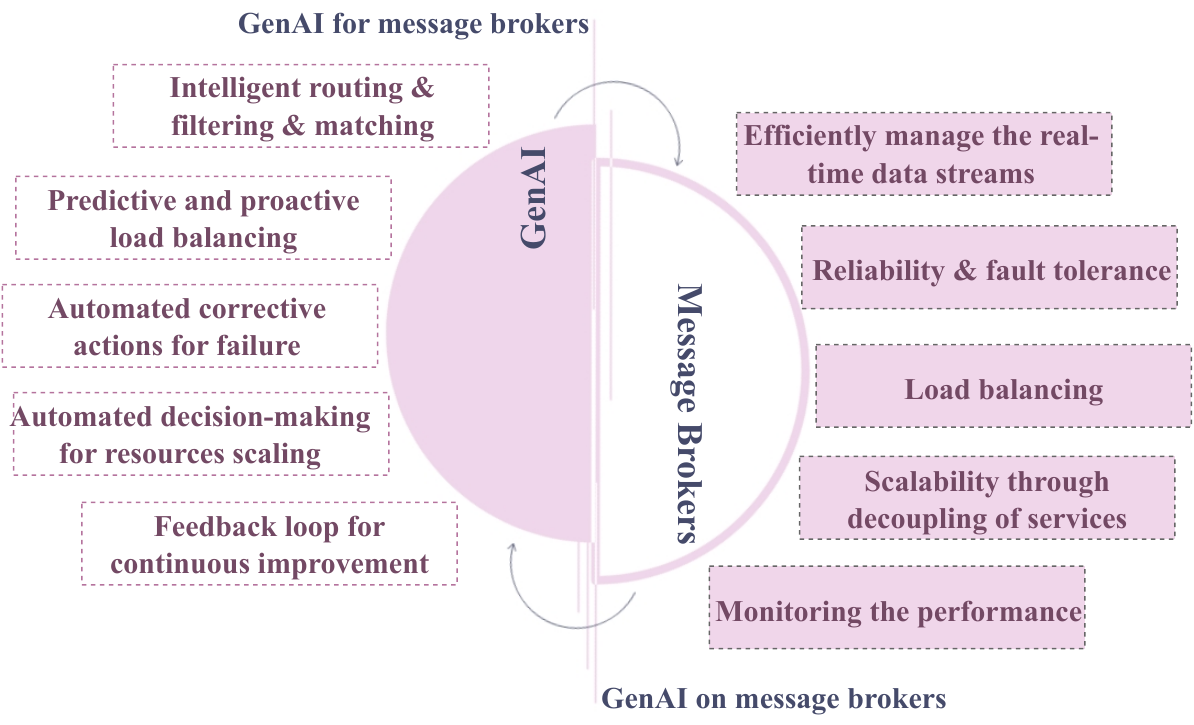}
\caption{Benefits provided by integrating a message broker with a GenAI agent.}\label{fig:integration with GenAI}
\end{figure*}

Following milestone studies in edge intelligence~\cite{park2019wireless,loven2019edgeai,deng2020edge}, the interaction between brokers and GenAI can be separated into two different categories: the benefits GenAI can bring for message brokers, and the benefits message brokers can bring for GenAI.
\cref{fig:integration with GenAI} further illustrates the expected benefits of integrating a message broker with GenAI.

\subsection{GenAI for message brokers}



GenAI has the potential to complement and enhance the intelligence and efficiency of message brokers, particularly by supporting the prediction of routing decisions to avoid busy routes and by enabling more nuanced filtering of messages according to specific criteria~\cite{dhoni2023exploring}. When integrated with pre-GenAI machine learning techniques, GenAI’s predictive capabilities~\cite{dhoni2023exploring} could assist in the proactive distribution of workloads or services across various nodes, potentially minimizing access times.

Furthermore, GenAI's ability to identify patterns that may induce errors~\cite{dhoni2023exploring} offers a promising avenue for augmenting message brokers with automatic corrective measures, potentially increasing system reliability and reducing downtime. Similarly, GenAI could assist in determining the optimal times for scaling resources, ensuring more efficient utilization~\cite{dhoni2023exploring}.

By leveraging GenAI’s advanced capabilities in understanding, interpreting, and generating text~\cite{chiu2023impact}, there is an opportunity to improve topic matching accuracy by analyzing the content of messages, enhancing the precision with which messages are delivered to their intended recipients. Additionally, GenAI’s capacity for learning from ongoing interactions and its explainability could create a continuous feedback loop~\cite{openai2023} that, when used alongside existing machine learning models, refines system performance over time.

\begin{figure*}
\centering
\includegraphics[width=1\textwidth]{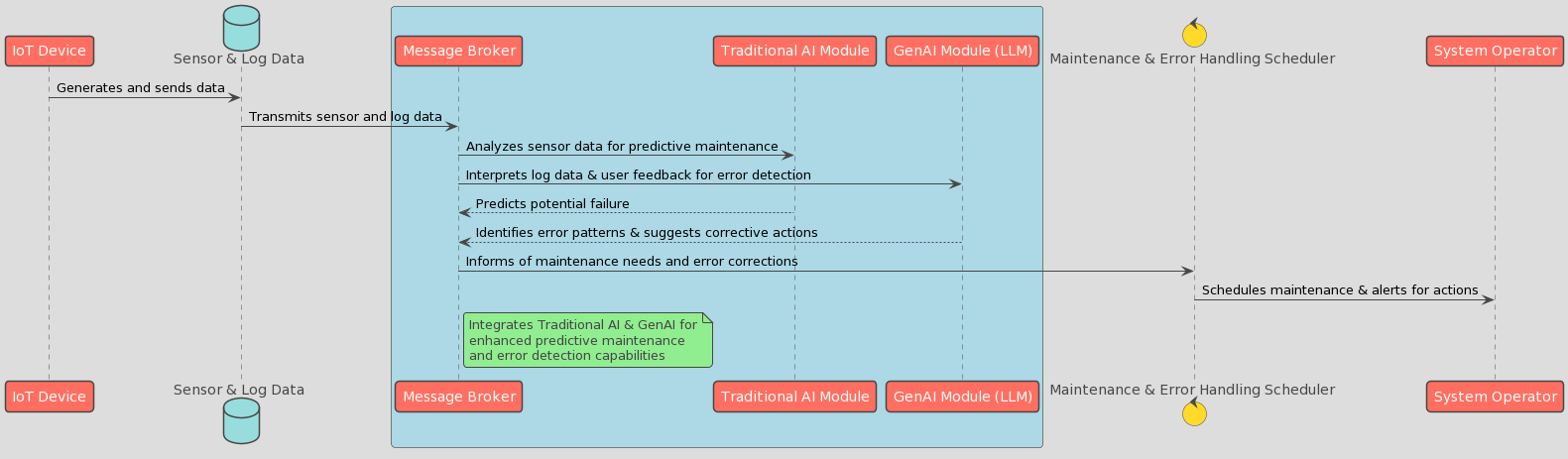}
\caption{GenAI for Message Brokers. This diagram illustrates how GenAI enhances traditional AI capabilities within a message broker system, optimizing routing decisions, error detection, and resource scaling based on real-time data analysis and predictive insights.}\label{fig:GenAIforMB}
\end{figure*}

Consider an IoT ecosystem with a message broker responsible for managing communications between thousands of devices across a smart city infrastructure. Traditional AI modules within the broker analyze sensor data to facilitate basic routing and load balancing. However, with the integration of GenAI, the system gains the ability to process and interpret large volumes of unstructured log data and textual feedback from devices and users in real-time. For instance, GenAI models, trained on vast datasets, can predict traffic congestion on network pathways by analyzing patterns from historical data and real-time environmental information \cite{celik2024dawn}. This predictive insight allows the message broker to reroute data flows dynamically, avoiding congested network nodes and reducing latency.

Moreover, GenAI's pattern recognition capabilities \cite{jin2023large} extend to identifying subtle signs of potential system failures or security breaches before they escalate. By analyzing error logs and user reports, GenAI can pinpoint anomalies that traditional systems might overlook, enabling preemptive maintenance and strengthening the network's security layout.

In the context of resource scaling, LLMs can be used for time series analysis to intercept trends in data traffic and device engagement to forecast demand spikes \cite{zhang2024large}. This foresight enables the system to scale resources up or down efficiently, ensuring optimal performance without wastage of bandwidth or computing power. The continuous learning aspect of GenAI, fueled by an ongoing feedback loop, ensures that the message broker's performance and decision-making processes improve over time, adapting to the evolving needs of the smart city infrastructure.

This example showcases the  potential of integrating GenAI with traditional AI in message brokers, significantly enhancing operational efficiency and system responsiveness in complex environments. Figure \ref{fig:GenAIforMB} illustrates this integration, highlighting the complementary roles of GenAI and traditional AI in optimizing the message brokering process.

\subsection{GenAI on message brokers}


Message brokers, leveraging asynchronous communication capabilities~\cite{ibmmq9}, can operate to enhance GenAI task processing efficiency. This approach facilitates interactions that do not necessitate real-time communication, streamlining the processing of real-time data streams. By decoupling the sending and receiving processes, message brokers can offer a flexible and scalable solution for managing the complex data workflows associated with GenAI applications.

Central to message brokers is their robust mechanism to prevent message loss~\cite{Oracle4}, which is key in conserving computational resources and safeguarding critical data for GenAI applications. These systems are equipped with tools designed to monitor and regulate message flow effectively. This functionality is fundamental in preserving the performance and operational integrity of GenAI by ensuring that data is processed efficiently and reliably, mitigating the risk of bottlenecks and data overflow.

Furthermore, message brokers facilitate service decoupling \cite{ibmmq9}, enabling GenAI to manage growing workloads more effectively. The broker's ability to distribute tasks to different nodes allows GenAI to achieve a balanced load distribution~\cite{ibmmq9}, with different components running on different nodes. This is particularly important for GenAI deployments in the computing continuum, providing them with access to local environments with limited computational capacity. In such cases, a message broker can act as an intermediary, bridging sensors with the perception module and conveying actions to the actuators and responses to the user~\cite{tarkoma2023ai,loven2023can}. 

Within the critical \textit{brain component}, 
message brokers play a vital role in linking various sub-components, each offering distinct features or possessing heterogeneous computational resources and functionalities.
By promoting interoperability and collaboration among diverse LLMs that contribute to decision-making, message brokers can significantly boost the performance of agents on complex tasks. This collaborative framework also facilitates the balanced offloading of tasks, optimizing the utilization of computing, communication, and storage resources across the network.



However, integrating GenAI into the pub/sub form requires careful planning and customization to mitigate the risks of biased or misleading outcomes.
Addressing hallucination problems \cite{rawte2023survey}, privacy \cite{gupta2023chatgpt}, and ethical considerations \cite{bang2023examination} is paramount to align GenAI implementations with specific system objectives~\cite{10198233}. Consequently, message brokers must adopt advanced methodologies to manage the extensive data volumes generated and consumed by GenAI applications effectively. Implementing computing- and networking-aware orchestration and resource management techniques represents a key strategy in this context. These approaches are not only essential for enhancing connectivity but also play a crucial role in reducing the computational demands and latency that are often associated with GenAI tasks.


In practical terms, the application of model compression methods and model training acceleration techniques can significantly mitigate GenAI's computational requirements and delays, addressing one of the major challenges in deploying these advanced systems efficiently. Furthermore, an efficient resource management process that enables workload distribution based on the capacity of each node directly contributes to improving the performance of the distributed system. This approach ensures that computational and storage resources are utilized optimally, preventing bottlenecks and ensuring smooth operation even under varying loads.

Moreover, transforming GenAI input and output data into priority-based smart data facilitates more timely and effective processing.  
Embedding a prioritization mechanism similar to the QoS levels defined in MQTT \cite{naik2017choice} serves as a suitable approach, ensuring that critical tasks are processed with the urgency they require. This adaptation enhances the responsiveness of GenAI systems by ensuring that high-priority data is attended to promptly, mirroring the efficiency and reliability seen in established communication protocols.

\begin{figure*}[ht]
\centering
\includegraphics[width=1\textwidth]{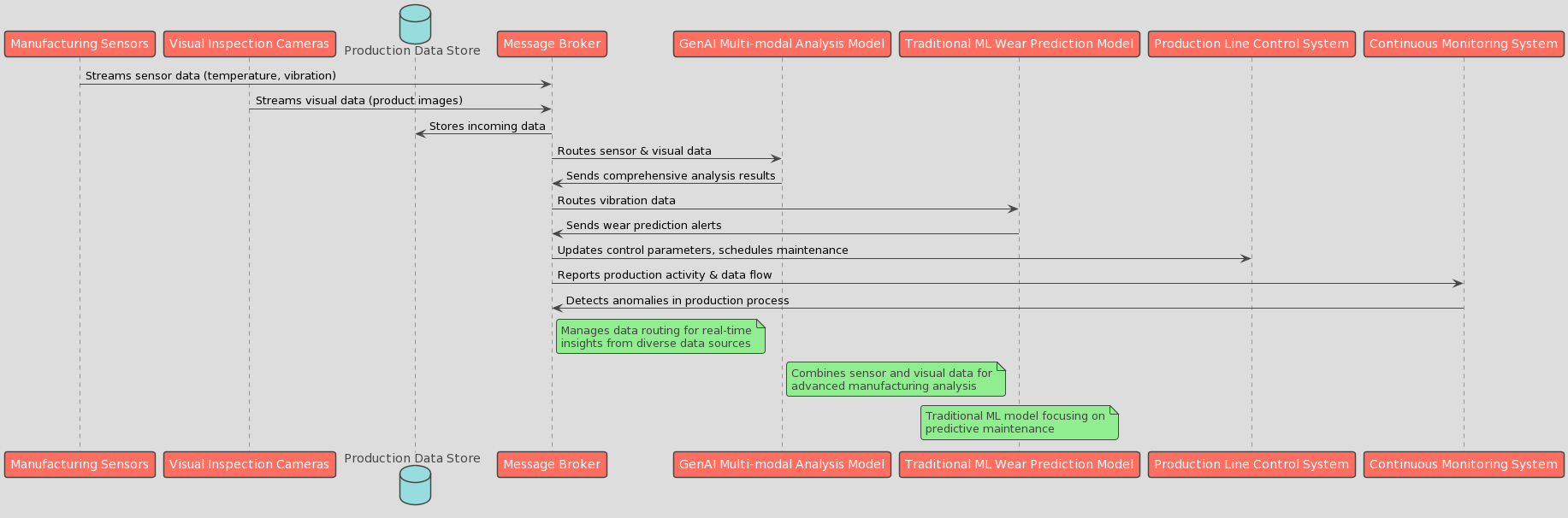}
\caption{Message brokers for GenAI. This diagram illustrates a smart manufacturing system where a message broker facilitates the flow of multimodal data from sensors and cameras to a GenAI model and a traditional ML model. The GenAI model performs comprehensive analysis by combining diverse data types, while the ML model focuses on predictive maintenance tasks.}\label{fig:MBforGenAI}
\end{figure*}

Additionally, the selection of the right models and their continuous adaptation in response to evolving data landscapes are crucial for maintaining the accuracy and relevance of GenAI outputs. In this context, message brokers play a pivotal role by incorporating capabilities for model fine-tuning \cite{min2023recent}, continual and in-context learning. These features enable dynamic adjustments to the models based on real-time data, ensuring that the GenAI system remains effective and up-to-date. This requirement underscores the necessity for message brokers to support not just the routing and handling of messages, but also the intelligent adaptation of GenAI models to changing conditions, thereby maximizing the potential of GenAI applications in diverse environments. In this respect, a continuous monitoring system that can promptly detect anomalies and data loss becomes fundamental. This system enables corrections and adaptations in real time, fostering the necessary model adaptation and ensuring that the GenAI systems remain both robust and responsive to the dynamic nature of real-world data and application demands.

We have outlined several key features on how message brokers can facilitate the integration and management of GenAI applications within various domains. For instance, Figure \ref{fig:MBforGenAI} demonstrates a practical application in smart manufacturing, where a message broker orchestrates the flow of multimodal data to both GenAI and traditional ML models. This setup enhances real-time decision-making and operational efficiency, showcasing the dynamic capabilities of message brokers in supporting advanced analytics.

In the following sections, we will discuss advanced methods designed to enhance the functionality of message brokers within the GenAI context, overcoming deployment challenges~\cite{tarkoma2023ai,loven2023can}. By tackling these issues, we aim for the seamless integration and optimal performance of GenAI applications. This exploration will include discussions on how specific platforms and frameworks can be leveraged to enable our envisioned approach, highlighting their potential adaptations in the analyzed context. 

\subsection{Semantic Communication}\label{sec:semantic}

Effective management and real-time analysis of vast data volumes are crucial for the development of GenAI applications. This necessitates a flexible system capable of accommodating a variety of data types with precision. Traditional message broker systems, while efficient in basic data routing, often struggle to cope with the complexity and volume of data typical in GenAI environments, limiting their effectiveness in scenarios requiring nuanced understanding and processing of data content. To address these limitations and reduce the strain on communication networks, embedding efficient communication mechanisms, such as semantic communication \cite{xia2023generative,jiang2023semantic}, within message brokers is essential. This integration, particularly leveraging the capabilities of LLMs, can enable intelligent, automated feature selection that aligns with subscriber needs, enhancing the broker's ability to manage data-rich content more effectively \cite{donta2023governance}. While this approach may not directly minimize latency due to the inference time required by LLMs, it significantly improves the efficiency of data search, match, and mapping processes, thereby optimizing overall system performance in handling and distributing relevant information.

\begin{figure*}
\centering
\includegraphics[width=1\textwidth]{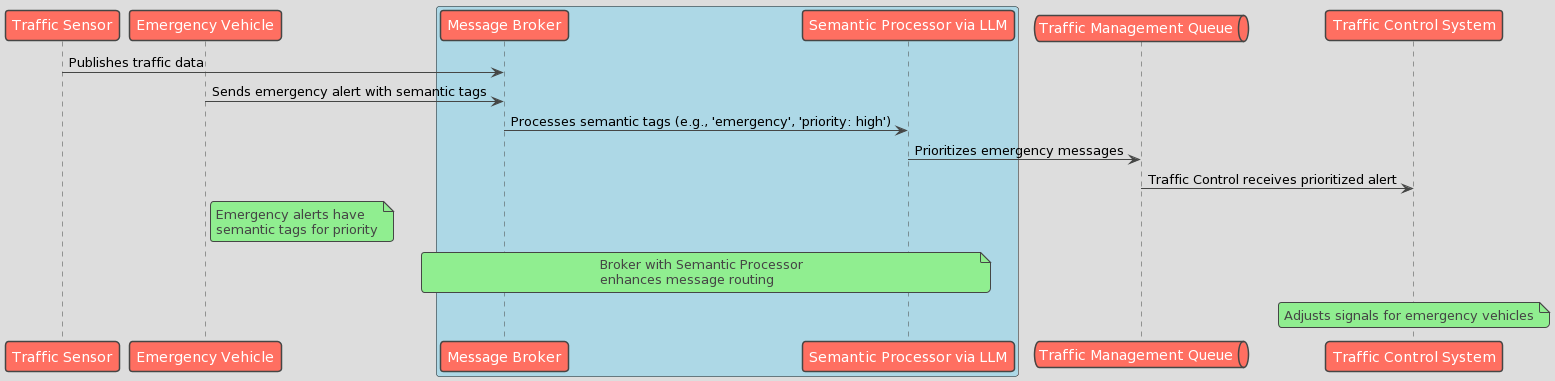}
\caption{GenAI-enabled and Semantic Communication. This diagram illustrates a traffic management system where a message broker, enhanced with semantic processing and LLMs capabilities, prioritizes and analyzes traffic data and emergency alerts.}\label{fig:SemanticComm}
\end{figure*}

Following this, message brokers equipped with dynamic prioritization capabilities can intelligently identify and route high-priority messages by incorporating semantic communication technology. This prioritization allows for the handling of messages based not just on the criteria within the message header, but also on the content itself, enhancing the relevance and timeliness of information delivery. Although integrating semantic communication with a broker introduces demands for high scalability, processing power, and memory to manage large datasets effectively, it is a crucial step towards mitigating network congestion and optimizing the use of network resources. Moreover, this sophisticated processing capability must be balanced with robust security features to ensure sensitive data is handled securely, highlighting the need for a comprehensive approach to upgrade message broker systems for the GenAI era.

In delay-sensitive applications like healthcare, this integration is vital to assigning priorities based on its deep understanding of data patterns and sensitivities and its subscribers’ specific needs. For example, in the healthcare scenario, GenAI can analyze large amounts of medical data to identify urgent cases, alerting healthcare professionals of critical patient needs or alarming health trends~\cite{oniani2023military}. 
In a similar fashion, the effective sinergy of LLMs with semantic processing capabilities within message brokers can also be observed in smart city traffic management (Fig.\ref{fig:SemanticComm}). Here, the combination of semantic tags from emergency vehicles and real-time traffic sensor data, when processed through an advanced LLM, enables the system to prioritize and analyze critical information promptly. This GenAI-enhanced approach not only interprets the urgency and context of incoming data but also predicts and optimizes traffic flow in response to dynamic urban conditions. By doing so, it ensures that emergency responses are effective, minimizing delays and improving public safety. Building upon the foundation laid by traditional ML techniques, GenAI complements these approaches by incorporating advanced natural language understanding and context-aware processing capabilities. This allows for a more peculiar analysis and interpretation of complex data sets, enhancing the system's ability to make informed decisions rapidly and accurately.

As examples, integration of semantic communication with RabbitMQ improves message routing capabilities through embedding semantic annotations into messages. As a result of these annotations, the broker can understand the meaning and context of each message, enhances message routing and efficient delivery. 

Furthermore, Apache RocketMQ, with its support for message broadcasting and filtering~\cite{ApacheRocketMQ}, has the potential to offer advanced message routing and filtering capabilities by leveraging semantic communication. These capabilities involve interpreting messages' semantic content, allowing for more precise and context-sensitive handling. 

Finally, Anypoint MQ's intelligent routing~\cite{AnypointMQ} combined with semantic filtering allows messages to be identified according to their inherent importance and meaning. It also ensures faster processing, routing of urgent and high priority content, and timely responses.

\subsection{Dynamic Data and Model Management}\label{sec:model}
GenAI-based applications require a data and model management system that not only simplifies the construction of AI models but also optimizes efficiency and effectiveness. Such a system should minimize the need for human intervention in selecting ML models, enhancing real-time responsiveness with high model's accuracy, and improving real-time inference capabilities when integrated with message brokers. LLMs can expedite the model selection process and boost the deployment efficiency and precision of AI solutions~\cite{zhang2023mlcopilot,shen2023large}.

Moreover, integrating GenAI models with message brokers involves managing and directing the related data flows. Since processing GenAI models is computationally intensive, with large amounts of real-time data, it needs to be scalable and distributed to cope with the varying workloads. Such integration also must also consider the handling of streaming data. 

As an example, due to its distributed streaming capabilities, Apache Pulsar~\cite{Pulsar} can be integrated with GenAI models that often require handling large datasets for training and inference. Brokers also need to handle a variety of data formats for integration with GenAI models. Red Hat AMQ, supporting multiple message patterns~\cite{RedHatAMQ}, could ensure compatibility with diverse GenAI data types. 

Because GenAI models such as LLMs often process sensitive or personal data, robust security measures are critical. For example, Google Cloud Pub/Sub~\cite{GoogleCloudPub/Sub} incorporates multiple integrated security measures to protect confidential data being transmitted. These measures include authentication and encryption using keys managed by Google, along with the latest Data Encryption Key (DEK) technology, among others~\cite{GoogleCloudPub/Sub}.

Moreover, Amazon Kinesis~\cite{Kinesis} offers  reliable data transmission methods, utilizing synchronous data replication, checkpointing mechanisms, and error handling strategies~\cite{Kinesis,Kinesis3}. 

Further, Apache Kafka's~\cite{Kafka} robust architecture, equipped with a stream processing feature, a cluster of brokers, and topic partition features, enables it to efficiently handle parallel processing tasks and feed the data to GenAI models. 

\subsection{Training Acceleration}\label{sec:accel}

The training of GenAI models requires a significant amount of computation and time. By incorporating  training acceleration methods~\cite{li2023colossal} into a message broker system, training time can be reduced, computational resources saved, and models deployed more rapidly.

The most common technique is sequence parallelism (SP), in which the input sequence is divided into smaller sub-sequences and processed in parallel~\cite{li2021sequence}. Another method involves selective activation re-computation, in which only the necessary parts of the model are recalculated during training, rather than the entire model~\cite{korthikanti2023reducing}. A fine-tuning technique involves adjusting the parameters of an existing model rather than training a new one from scratch. As a result, training data and computations can be reduced \cite{radford2018improving}.

Furthermore, tensor parallelism works by splitting the model across multiple GPUs and processing different parts of the model in parallel~\cite{li2023colossal}. The mixed-precision training technique is one of the most effective ways to speed up training on modern GPUs~\cite{micikevicius2018mixed}. In this technique, the amount of memory required during training is reduced by using lower-precision data types.

As of our knowledge, no broker currently supports model training acceleration. However, a message broker offers a variety of useful functionalities, including the scheduling and distribution of data compression tasks, as well as monitoring the performance of compressed models. These methods are discussed in more detail in \cref{sec:orch,sec:monitor}.


\subsection{Dynamic Model Compression}\label{sec:comp}

Integrating GenAI models on resource-constrained nodes in the computing continuum often requires model compression, especially with high-dimensional models, strenuous computational tasks, and low latency requirements~\cite{bariah2023large}. Among the compression strategies, pruning is prominent, involving the removal of superfluous elements from a model to decrease its size and complexity~\cite{jiang2022model}. Another technique is Knowledge Distillation, where a smaller 'student' model learns to replicate the functionality of a larger 'teacher' model~\cite{zhu2023survey, polino2018model}.

Furthermore, various quantization methods are utilized to decrease model parameter precision, thereby cutting memory usage and computational needs substantially~\cite{polino2018model}. Another technique is low-rank factorization, which simplifies weight matrices with lower-rank approximations to reduce model size and computational demands~\cite{hsu2022language}.

Currently, no broker supports model compression as such. However, message brokers can provide essential capabilities such as scheduling and distributing compression tasks, or monitoring the performance of compressed models. These methods are discussed in more detail in  \cref{sec:orch,sec:monitor}.


\subsection{Dynamic Orchestration}\label{sec:orch}
Integration with GenAI requires effective resource management. The broker must allocate and manage resources such as CPU, memory, and storage to handle computational requirements. Furthermore, supporting feedback mechanisms is essential to allow continuous learning and improvement of the system based on the output and performance of the GenAI agent.
However, the process of efficient resource usage and allocation requires parallel processing capability. 

For instance, Celery can handle, maintain, and schedule distributed tasks across various nodes, thereby enhancing the efficiency of GenAI agents~\cite{Celery}. Apache ActiveMQ provides efficient resource allocation~\cite{ApacheActiveMQ}. Thus, optimizing Apache ActiveMQ is critical to providing intelligent resource allocation to ensure balanced workloads across different nodes. 

Amazon SQS provides a reliable queue service for handling messages that facilitates microservices and distributed systems decoupling~\cite{SQS}. Further, the robust architecture of Apache Kafka, which uses clusters of brokers to handle data distribution and partition topics for scalability, makes it excellent for supporting distributed event streaming~\cite{Kafka}. Additionally, Azure Service Bus provides advanced scheduling features and message orchestration in distributed environments~\cite{AzureServiceBus5,AzureServiceBus4}.

\subsection{AIOps/MLOps and Monitoring}\label{sec:monitor}

MLOps, merging DevOps principles with ML, is central to the advancement of ML and AI, streamlining the lifecycle of models from development to deployment. A critical feature within this field is the monitoring of deployed models, crucial for the uninterrupted and reliable operation of message broker systems. This practice enables real-time insights into model metrics, resource usage, and system irregularities, creating a proactive environment for identifying and addressing issues promptly. Furthermore, MLOps facilitates the setting up of automated alerts and triggers, enhancing responsiveness to anomalies and minimizing downtime~\cite{loven2023can,kreuzberger2023machine}.

A Continuous Diagnostics and Mitigation (CDM) program plays a vital role in network security by analyzing network behavior and thwarting unauthorized access, thereby enabling prompt responses and maintaining network integrity. Beyond autonomous configuration management and monitoring device availability, CDM programs conduct continuous health assessments of devices and evaluate their environmental footprint. This continuous surveillance helps identify potential threats, bolstering processes to enhance security measures. Furthermore, CDM ensures the protection of sensitive information against unauthorized access or breaches~\cite{Praveen2023exploring}.

However, frameworks such as MLOps and CDM with are resource-hungry. Integrating them with message brokers requires careful consideration of, for example, the computation capacity available locally, as well as the distribution of the related tasks, to avoid the starvation of regular operations. 


As KubeMQ supports multistage pipelines~\cite{KubeMQ}, it could be essential for managing data pipelines within MLOps frameworks. Moreover, HiveMQ and Amazon Kinesis can contribute to MLOps integration by facilitating real-time monitoring and alert systems integration, since they support real-time device monitoring~\cite{HiveMQ,Kinesis}.

Finally, tracking GenAI model performance can be effectively achieved through integration with IBM MQ, which has robust monitoring and tracing capabilities~\cite{ibmmq}. Solace PubSub could potentially assist in identifying and resolving issues related to message routing, delivery, and processing~\cite{Solace1}, thus facilitating control and modeling activities in MLOps environments.



\subsection{Summary of message broker enhancement methods}
While we have thoroughly explored the possible interplay between existing message broker technologies and their specific features to meet GenAI requirements, it is important to emphasize that our assessment of the suitability of a certain technology for specific tasks is indicative of their potential in the given context. Selecting any technology must be informed by a comprehensive analysis of the application and infrastructure topology requirements, data handling needs, and the particular features of these tools that align with the envisioned objectives. Adopting this approach can guarantee that the chosen solution not only fulfills immediate operational demands but also possesses the necessary scalability and flexibility for future growth and increased complexity. This consideration is crucial as we move towards the conclusion of our discussion, underscoring the importance of strategic technology selection in the dynamic landscape of GenAI-enhanced communication systems. The challenges, opportunities, and our strategic view on utilizing these technologies, along with the related subsections, are summarized in \cref{tab:opportunities}.

\begin{table*}[]
    \centering
    \caption{Opportunities \& Challenges for enhancing message brokers' functionality within GenAI}
    \resizebox{!}{.1\paperheight}{%
    
    \begin{tabular}{lcl}
\hline
\rowcolor[HTML]{F4EFEF} 
\multicolumn{1}{|c|}{\cellcolor[HTML]{F4EFEF}\textbf{Opportunities}} &
  \multicolumn{1}{c|}{\cellcolor[HTML]{F4EFEF}\textbf{\begin{tabular}[c]{@{}c@{}}Correspondong\\ Section\end{tabular}}} &
  \multicolumn{1}{c|}{\cellcolor[HTML]{F4EFEF}\textbf{Challenges}} \\ \hline
\rowcolor[HTML]{FFFFFF} 
\multicolumn{1}{|l|}{\cellcolor[HTML]{FFFFFF}\begin{tabular}[c]{@{}l@{}}\textbf{Semantic Communication} to reduce communication\\ networks strain and streamline the data-rich content\\ transmission.\end{tabular}} &
  \multicolumn{1}{c|}{\cellcolor[HTML]{FFFFFF}\cref{sec:semantic}} &
  \multicolumn{1}{l|}{\cellcolor[HTML]{FFFFFF}\begin{tabular}[c]{@{}l@{}}Computationally intensive to do locally.\\ Processing sensitive data requires robust security measures.\end{tabular}} \\ \hline
\rowcolor[HTML]{FFFFFF} 
 &
  \multicolumn{1}{l}{\cellcolor[HTML]{FFFFFF}} &
   \\ \hline
\rowcolor[HTML]{FFFFFF} 
\multicolumn{1}{|l|}{\cellcolor[HTML]{FFFFFF}\begin{tabular}[c]{@{}l@{}}\textbf{Dynamic Data and Model Management} to minimize\\ the need for human intervention in selecting ML models\\ and enhance real-time responsiveness accuracy.\end{tabular}} &
  \multicolumn{1}{c|}{\cellcolor[HTML]{FFFFFF}\cref{sec:model}} &
  \multicolumn{1}{l|}{\cellcolor[HTML]{FFFFFF}\begin{tabular}[c]{@{}l@{}}Computationally intensive to do locally.\\ May require coordination between clients and broker to exchange \\information on, e.g., model architectures.\end{tabular}} \\ \hline
\rowcolor[HTML]{FFFFFF} 
 &
  \multicolumn{1}{l}{\cellcolor[HTML]{FFFFFF}} &
   \\ \hline
\rowcolor[HTML]{FFFFFF} 
\multicolumn{1}{|l|}{\cellcolor[HTML]{FFFFFF}\begin{tabular}[c]{@{}l@{}}\textbf{Training Acceleration} to reduce training time, save \\ computational resources, and rapidly deploy models.\end{tabular}} &
  \multicolumn{1}{c|}{\cellcolor[HTML]{FFFFFF}\cref{sec:accel}} &
  \multicolumn{1}{l|}{\cellcolor[HTML]{FFFFFF}\begin{tabular}[c]{@{}l@{}}Requires ability to manage load balancing \& data dependencies.\\ Compatibility with hardware configuration.\end{tabular}} \\ \hline
\rowcolor[HTML]{FFFFFF} 
 &
  \multicolumn{1}{l}{\cellcolor[HTML]{FFFFFF}} &
   \\ \hline
\rowcolor[HTML]{FFFFFF} 
\multicolumn{1}{|l|}{\cellcolor[HTML]{FFFFFF}\begin{tabular}[c]{@{}l@{}}\textbf{Dynamic Model Compression} to save resources and\\ improve response time.\end{tabular}} &
  \multicolumn{1}{c|}{\cellcolor[HTML]{FFFFFF}\cref{sec:comp}} &
  \multicolumn{1}{l|}{\cellcolor[HTML]{FFFFFF}\begin{tabular}[c]{@{}l@{}}Computationally intensive to do locally.\\ May require Coordination between clients and broker to exchange \\information on, e.g., model architectures.\end{tabular}} \\ \hline
\rowcolor[HTML]{FFFFFF} 
 &
  \multicolumn{1}{l}{\cellcolor[HTML]{FFFFFF}} &
   \\ \hline
\rowcolor[HTML]{FFFFFF} 
\multicolumn{1}{|l|}{\cellcolor[HTML]{FFFFFF}\begin{tabular}[c]{@{}l@{}}\textbf{Dynamic Orchestration} to optimize use\\ of resources.\end{tabular}} &
  \multicolumn{1}{c|}{\cellcolor[HTML]{FFFFFF}\cref{sec:orch}} &
  \multicolumn{1}{l|}{\cellcolor[HTML]{FFFFFF}\begin{tabular}[c]{@{}l@{}}Requires novel, highly distributed, efficient, and secure resource \\orchestration methods.\end{tabular}} \\ \hline
\rowcolor[HTML]{FFFFFF} 
 &
  \multicolumn{1}{l}{\cellcolor[HTML]{FFFFFF}} &
   \\ \hline
\rowcolor[HTML]{FFFFFF} 
\multicolumn{1}{|l|}{\cellcolor[HTML]{FFFFFF}\begin{tabular}[c]{@{}l@{}}\textbf{AIOps/MLOps and Monitoring} to enhance responsiveness\\ to anomalies and minimize downtime.\end{tabular}} &
  \multicolumn{1}{c|}{\cellcolor[HTML]{FFFFFF}\cref{sec:monitor}} &
  \multicolumn{1}{l|}{\cellcolor[HTML]{FFFFFF}\begin{tabular}[c]{@{}l@{}}Computationally intensive to do locally.\\ Efficient and dynamic prioritization between monitoring and \\regular tasks.\end{tabular}} \\ \hline
\end{tabular}%
}
    
    \label{tab:opportunities}
\end{table*}

\section{Conclusion} \label{sec:conclusion}

In this paper, we have provided a comprehensive overview of contemporary message brokers, delineating their features, capabilities, and limitations with an eye toward their application within GenAI frameworks. Our analysis spanned a broad spectrum of criteria and we delved into the inherent limitations of existing message brokers when confronted with the demands of GenAI applications, prompting a reflection on the essential attributes of an ideal message broker framework designed to seamlessly integrate with GenAI technologies. In addressing these challenges, we analyzed several requirements to be satisfied in order to bolstering the efficacy of message brokers in facilitating the rapid evolution and deployment of GenAI applications. 

Through a comprehensive analysis of the current state, challenges, and forward-looking strategies for message brokers, this study lays the groundwork for the development of more adaptable and efficient GenAI-enabled communication systems. Such systems are envisioned to not only distribute data with increasing efficiency but also to ensure the delivery of high-quality service, manage resources with greater intelligence, and satisfy the increasing demands of GenAI applications.

Finally, our exploration underscores the critical need for message brokers to evolve in tandem with technological advancements and GenAI requirements. By identifying opportunities for improvement, this paper aims to boost further research and development efforts focused on creating message broker frameworks that are not only robust and scalable but also closely aligned to the peculiarities of GenAI-driven data communication.

\section*{Acknowledgement}
This research is supported by the Research Council of Finland (former Academy of Finland) 6G Flagship Program (Grant Number: 346208), and by Business Finland through the Neural pub/sub research project (diary number 8754/31/2022).

\bibliographystyle{IEEEtran}
\bibliography{ref}

\begin{thebibliography}{100}
\providecommand{\url}[1]{#1}
\csname url@samestyle\endcsname
\providecommand{\newblock}{\relax}
\providecommand{\bibinfo}[2]{#2}
\providecommand{\BIBentrySTDinterwordspacing}{\spaceskip=0pt\relax}
\providecommand{\BIBentryALTinterwordstretchfactor}{4}
\providecommand{\BIBentryALTinterwordspacing}{\spaceskip=\fontdimen2\font plus
\BIBentryALTinterwordstretchfactor\fontdimen3\font minus \fontdimen4\font\relax}
\providecommand{\BIBforeignlanguage}[2]{{%
\expandafter\ifx\csname l@#1\endcsname\relax
\typeout{** WARNING: IEEEtran.bst: No hyphenation pattern has been}%
\typeout{** loaded for the language `#1'. Using the pattern for}%
\typeout{** the default language instead.}%
\else
\language=\csname l@#1\endcsname
\fi
#2}}
\providecommand{\BIBdecl}{\relax}
\BIBdecl

\bibitem{chatgpt}
OpenAI, ``Chatgpt,'' \url{https://chat.openai.com/}, last accessed: \today.

\bibitem{shen2024large}
Y.~Shen, J.~Shao, X.~Zhang, Z.~Lin, H.~Pan, D.~Li, J.~Zhang, and K.~B. Letaief, ``Large language models empowered autonomous edge ai for connected intelligence,'' \emph{IEEE Communications Magazine}, 2024.

\bibitem{tarkoma2023ai}
S.~Tarkoma, R.~Morabito, and J.~Sauvola, ``Ai-native interconnect framework for integration of large language model technologies in 6g systems,'' \emph{arXiv preprint arXiv:2311.05842}, 2023.

\bibitem{bariah2023large}
L.~Bariah, Q.~Zhao, H.~Zou, Y.~Tian, F.~Bader, and M.~Debbah, ``Large language models for telecom: The next big thing?'' \emph{arXiv preprint arXiv:2306.10249}, 2023.

\bibitem{kokkonen2022autonomy}
H.~Kokkonen, L.~Lov{\'e}n, N.~H. Motlagh, A.~Kumar, J.~Partala, T.~Nguyen, V.~C. Pujol, P.~Kostakos, T.~Lepp{\"a}nen, A.~Gonz{\'a}lez-Gil \emph{et~al.}, ``Autonomy and intelligence in the computing continuum: Challenges, enablers, and future directions for orchestration,'' \emph{arXiv preprint arXiv:2205.01423}, 2022.

\bibitem{motlagh2022edge}
N.~H. Motlagh, L.~Lov{\'e}n, J.~Cao, X.~Liu, P.~Nurmi, S.~Dustdar, S.~Tarkoma, and X.~Su, ``Edge computing: The computing infrastructure for the smart megacities of the future,'' \emph{Computer}, vol.~55, no.~12, pp. 54--64, 2022.

\bibitem{wang2023overview}
Y.-C. Wang, J.~Xue, C.~Wei, and C.-C.~J. Kuo, ``An overview on generative ai at scale with edge-cloud computing,'' 2023.

\bibitem{dustdar2022distributed}
S.~Dustdar, V.~C. Pujol, and P.~K. Donta, ``On distributed computing continuum systems,'' \emph{IEEE Transactions on Knowledge and Data Engineering}, vol.~35, no.~4, pp. 4092--4105, 2022.

\bibitem{Praveen2023exploring}
D.~Praveen~Kumar, M.~Ilir, C.~P. Victor, S.~Boris, and D.~Schahram, ``Exploring the potential of distributed computing continuum systems,'' \emph{Computers}, 2023.

\bibitem{8782370}
F.~Ozturk and A.~M. Ozdemir, ``Content-based publish/subscribe communication model between iot devices in smart city environment,'' in \emph{2019 7th International Istanbul Smart Grids and Cities Congress and Fair (ICSG)}, 2019, pp. 189--193.

\bibitem{7098706}
R.~Wadhwa, A.~Mehra, P.~Singh, and M.~Singh, ``A pub/sub based architecture to support public healthcare data exchange,'' in \emph{2015 7th International Conference on Communication Systems and Networks (COMSNETS)}, 2015, pp. 1--6.

\bibitem{eugster2003many}
P.~T. Eugster, P.~A. Felber, R.~Guerraoui, and A.-M. Kermarrec, ``The many faces of publish/subscribe,'' \emph{ACM computing surveys (CSUR)}, vol.~35, no.~2, pp. 114--131, 2003.

\bibitem{tarkoma2012publish}
S.~Tarkoma, \emph{Publish/subscribe systems: design and principles}.\hskip 1em plus 0.5em minus 0.4em\relax John Wiley \& Sons, 2012.

\bibitem{donta2022survey}
P.~K. Donta, S.~N. Srirama, T.~Amgoth, and C.~S.~R. Annavarapu, ``Survey on recent advances in iot application layer protocols and machine learning scope for research directions,'' \emph{Digital Communications and Networks}, vol.~8, no.~5, pp. 727--744, 2022.

\bibitem{hasenburg2020managing}
J.~Hasenburg, F.~Stanek, F.~Tschorsch, and D.~Bermbach, ``Managing latency and excess data dissemination in fog-based publish/subscribe systems,'' in \emph{2020 IEEE international conference on fog computing (ICFC)}.\hskip 1em plus 0.5em minus 0.4em\relax IEEE, 2020, pp. 9--16.

\bibitem{redondi2019towards}
A.~E. Redondi, A.~Arcia-Moret, and P.~Manzoni, ``Towards a scaled iot pub/sub architecture for 5g networks: The case of multiaccess edge computing,'' in \emph{2019 IEEE 5th World Forum on Internet of Things (WF-IoT)}.\hskip 1em plus 0.5em minus 0.4em\relax IEEE, 2019, pp. 436--441.

\bibitem{loven2023can}
L.~Lov{\'e}n, R.~Morabito, A.~Kumar, S.~Pirttikangas, J.~Riekki, and S.~Tarkoma, ``How can ai be distributed in the computing continuum? introducing the neural pub/sub paradigm,'' \emph{arXiv preprint arXiv:2309.02058}, 2023.

\bibitem{chafi2022introduction}
F.~Z. Chafi, Y.~Fakhri, and F.~Z. A.~H. Aadi, ``Introduction to internet of things’ communication protocols,'' in \emph{Advanced Intelligent Systems for Sustainable Development (AI2SD’2020) Volume 2}.\hskip 1em plus 0.5em minus 0.4em\relax Springer, 2022, pp. 142--150.

\bibitem{maniezzo2023self}
V.~Maniezzo, M.~A. Boschetti, and P.~Manzoni, ``Self-adaptive publish/subscribe network design,'' in \emph{Metaheuristics: 14th International Conference, MIC 2022, Syracuse, Italy, July 11--14, 2022, Proceedings}.\hskip 1em plus 0.5em minus 0.4em\relax Springer, 2023, pp. 478--484.

\bibitem{pedrosa2021reducing}
F.~Pedrosa and L.~Rodrigues, ``Reducing the subscription latency in reliable causal publish-subscribe systems,'' in \emph{Proceedings of the 36th Annual ACM Symposium on Applied Computing}, 2021, pp. 203--212.

\bibitem{john2017survey}
V.~John and X.~Liu, ``A survey of distributed message broker queues,'' \emph{arXiv preprint arXiv:1704.00411}, 2017.

\bibitem{linthicum2000enterprise}
D.~S. Linthicum, \emph{Enterprise application integration}.\hskip 1em plus 0.5em minus 0.4em\relax Addison-Wesley Professional, 2000.

\bibitem{perry2001mqseries}
M.~Perry, C.~Delporte, F.~Demi, A.~Ghosh, and M.~Luong, \emph{MQSeries publish/subscribe applications}.\hskip 1em plus 0.5em minus 0.4em\relax IBM Redbooks, 2001.

\bibitem{hohpe2004enterprise}
G.~Hohpe and B.~Woolf, \emph{Enterprise integration patterns: Designing, building, and deploying messaging solutions}.\hskip 1em plus 0.5em minus 0.4em\relax Addison-Wesley Professional, 2004.

\bibitem{indrasiri2021design}
K.~Indrasiri and S.~Suhothayan, \emph{Design Patterns for Cloud Native Applications}.\hskip 1em plus 0.5em minus 0.4em\relax " O'Reilly Media, Inc.", 2021.

\bibitem{iakushkin2014messaging}
O.~Iakushkin and V.~Grishkin, ``Messaging middleware for cloud applications: Extending brokerless approach,'' in \emph{2014 2nd 2014 2nd International Conference on Emission Electronics (ICEE)}.\hskip 1em plus 0.5em minus 0.4em\relax IEEE, 2014, pp. 1--4.

\bibitem{ApacheActiveMQ}
T.~A.~S. Foundation, ``Apache activemq,'' \url{https://activemq.apache.org/}, last accessed: \today.

\bibitem{Fuse}
R.~Hat, ``Fuse message broker,'' \url{https://access.redhat.com /taxonomy/products/fuse-message-broker}, last accessed: \today.

\bibitem{ApacheQpid}
A.~S. Foundation, ``Apache qpid,'' \url{https://qpid.apache.org/}, last accessed: \today.

\bibitem{RabbitMQ}
R.~Technologies, ``Rabbitmq,'' \url{https://www.rabbitmq.com/}, last accessed: \today.

\bibitem{HornetQ}
R.~Hat, ``Hornetq,'' \url{https://hornetq.jboss.org/}, last accessed: \today.

\bibitem{RedHatAMQ}
I.~Red~Hat, ``Red hat amq,'' \url{https://www.redhat.com/en/technologies/jboss-middleware/amq}, last accessed: \today.

\bibitem{Celery}
C.~software, ``Celery,'' \url{https://docs.celeryq.dev/}, last accessed: \today.

\bibitem{JBossMessaging}
R.~Hat, ``Jboss messaging,'' \url{https://jbossmessaging.jboss.org/}, last accessed: \today.

\bibitem{OpenMQ}
Oracle, ``Openmq,'' \url{https://javaee.github.io/openmq/}, last accessed: \today.

\bibitem{Beanstalk}
I.~Philotic, ``Beanstalk,'' \url{https://beanstalkd.github.io/}, last accessed: \today.

\bibitem{Gearman}
Gearman, ``Gearman,'' \url{http://gearman.org/}, last accessed: \today.

\bibitem{Enduro/X}
Mavimax, ``Enduro/x,'' \url{https://www.endurox.org/}, last accessed: \today.

\bibitem{WSO2}
P.~Fremantle, ``Wso2,'' \url{https://wso2.com/}, last accessed: \today.

\bibitem{HiveMQ}
HiveMQ, ``Hivemq,'' \url{https://www.hivemq.com/}, last accessed: \today.

\bibitem{Redis}
R.~Labs, ``Redis,'' \url{https://redis.io/}, last accessed: \today.

\bibitem{EMQX}
E.~Technologies, ``Emqx,'' \url{https://www.emqx.io/}, last accessed: \today.

\bibitem{Pulsar}
T.~A.~S. Foundation, ``Apache pulsar,'' \url{https://pulsar.apache.org/}, last accessed: \today.

\bibitem{Kafka}
A.~S. Foundation, ``Apache kafka,'' \url{https://kafka.apache.org/}, last accessed: \today.

\bibitem{ApacheRocketMQ}
A.~Group, ``Apache rocketmq,'' \url{https://rocketmq.apache.org/}, last accessed: \today.

\bibitem{EclipseMosquitto}
Eclipse, ``Eclipse mosquitto,'' \url{https://mosquitto.org/}, last accessed: \today.

\bibitem{zeromq}
iMatix, ``Zero mq,'' \url{https://zeromq.org/}, last accessed: \today.

\bibitem{ApacheNiFi}
I.~Onyara, ``Apache nifi,'' \url{https://nifi.apache.org/}, last accessed: \today.

\bibitem{AblyRealtime}
A.~R. LTD, ``Ably realtime,'' \url{https://ably.com/}, last accessed: \today.

\bibitem{SamZa}
A.~S. Foundation, ``Apache samza,'' \url{https://samza.apache.org/}, last accessed: \today.

\bibitem{VerneMQ}
VerneMQ, ``Vernemq,'' \url{https://vernemq.com/}, last accessed: \today.

\bibitem{NServiceBus}
P.~Software, ``Nservicebus,'' \url{https://particular.net/nservicebus}, last accessed: \today.

\bibitem{kestrel}
Twitter, ``kestrel,'' \url{https://github.com/twitter-archive/kestrel}, last accessed: \today.

\bibitem{NSQ}
Bitly, ``Nsq,'' \url{https://nsq.io/}, last accessed: \today.

\bibitem{NATS}
S.~Communications, ``Nats,'' \url{https://nats.io/}, last accessed: \today.

\bibitem{KubeMQ}
KubeMQ, ``Kubemq,'' \url{https://kubemq.io/}, last accessed: \today.

\bibitem{ibmmq}
IBM, ``Ibm mq,'' \url{https://www.ibm.com/docs/en/ibm-mq}, last accessed: \today.

\bibitem{SQS}
A.~W.~S. (AWS), ``Amazon sqs,'' \url{https://aws.amazon.com/sqs/}, last accessed: \today.

\bibitem{MSMQ}
Microsoft, ``Msmq,'' \url{https://learn.microsoft.com/en-us/previous-versions/windows/desktop/msmq/}, last accessed: \today.

\bibitem{Oracle1}
Oracle, ``Oracle message broker,'' \url{https://docs.oracle.com/cd/E26576\_01/doc.312/e24948.pdf}, last accessed: \today.

\bibitem{TIBCORendezvous}
TIBCO, ``Tibco rendezvous,'' \url{https://www.tibco.com/products/tibco-rendezvous}, last accessed: \today.

\bibitem{TIBCOEnterpriseMessageService}
------, ``Tibco enterprise message service™,'' \url{https://www.tibco.com/products/tibco-enterprise-message-service}, last accessed: \today.

\bibitem{AnypointMQ}
MuleSoft, ``Anypoint mq,'' \url{https://docs.mulesoft.com/mq/}, last accessed: \today.

\bibitem{AzureServiceBus}
Microsoft, ``Azure service bus,'' \url{https://learn.microsoft.com/en-us/azure/service-bus-messaging/}, last accessed: \today.

\bibitem{SAP}
S.~AG, ``Sap netweaver process integration,'' \url{https://help.sap.com/docs/}, last accessed: \today.

\bibitem{Solace}
C.~Betts, ``Solace,'' \url{https://solace.com/}, last accessed: \today.

\bibitem{GoogleCloudPub/Sub}
Google, ``Google cloud pub/sub,'' \url{https://cloud.google.com/pubsub}, last accessed: \today.

\bibitem{AmazonMQ}
A.~W.~S. (AWS), ``Amazon mq,'' \url{https://aws.amazon.com/amazon-mq/}, last accessed: \today.

\bibitem{IntelMPILibrary}
Intel, ``Intel mpi library,'' \url{https://www.intel.com/content/ www/us/en/developer/tools/oneapi/mpi-library.html}, last accessed: \today.

\bibitem{Kinesis}
Amazon, ``Kinesis,'' \url{https://aws.amazon.com/kinesis/}, last accessed: \today.

\bibitem{AzureStorageQueue}
Microsoft, ``Azure storage queue,'' \url{https://learn.microsoft.com/en-us/azure/storage/queues/}, last accessed: \today.

\bibitem{IronMQ}
Iron.io, ``Ironmq,'' \url{http://www.iron.io/mq}, last accessed: \today.

\bibitem{Fuse1}
R.~Hat, ``Fuse message broker,'' \url{https://docs.huihoo.com/fuse/getting_started.pdf}, last accessed: \today.

\bibitem{ApacheQpid2}
A.~S. Foundation, ``Apache qpid,'' \url{https://people.apache.org/~jonathan/Programming-In-Apache-Qpid.html}, last accessed: \today.

\bibitem{RabbitMQ1}
R.~Technologies, ``Rabbitmq,'' \url{https://blog.rabbitmq.com/posts/2020/07/disaster-recovery-and-high-availability-101/}, last accessed: \today.

\bibitem{HornetQuser}
R.~Hat, ``Hornetq,'' \url{https://hornetq.sourceforge.net/docs/hornetq-2.1.2.Final/user-manual/en/html_single/index.html}, last accessed: \today.

\bibitem{HornetQprotocol}
------, ``Hornetq,'' \url{https://docs.jboss.org/hornetq/2.4.0.Final/docs/user-manual/html/interoperability.html}, last accessed: \today.

\bibitem{RedHatAMQ1}
I.~Red~Hat, ``Red hat amq,'' \url{https://access.redhat.com/documentation/}, last accessed: \today.

\bibitem{JBossCons}
R.~Hat, ``Jboss messaging,'' \url{https://docs.jboss.org/jbossmessaging/docs/usermanual-2.0.0.beta1/html}, last accessed: \today.

\bibitem{Beanstalkprotocol2}
I.~Philotic, ``Beanstalk,'' \url{https://raw.githubusercontent.com/beanstalkd/beanstalkd/master/doc/protocol.txt}, last accessed: \today.

\bibitem{Beanstalkprotocol}
------, ``Beanstalk,'' \url{https://github.com/beanstalkd/beanstalkd/wiki/Client-Libraries}, last accessed: \today.

\bibitem{Mavimax2022Enduro/X1}
Mavimax, ``Enduro/x,'' \url{https://github.com/endurox-dev/endurox/blob/master/doc/}, last accessed: \today.

\bibitem{WSO2memory}
P.~Fremantle, ``Wso2,'' \url{https://ei.docs.wso2.com/en/latest/micro-integrator/setup/performance_tuning/tuning_jvm_performance/}, last accessed: \today.

\bibitem{WSO2transport}
------, ``Wso2,'' \url{https://apim.docs.wso2.com/en/4.1.0/install-and-setup/setup/mi-setup/transport\_configurations/configuring-transports/}, last accessed: \today.

\bibitem{WSO2filter}
------, ``Wso2,'' \url{https://ei.docs.wso2.com/en/latest/micro-integrator/references/mediators/filter-Mediator/}, last accessed: \today.

\bibitem{HiveMQ2}
HiveMQ, ``Hivemq,'' \url{https://docs.hivemq.com/hivemq/}, last accessed: \today.

\bibitem{Redis2}
R.~Labs, ``Redis,'' \url{https://docs.redis.com/latest/rs/security/}, last accessed: \today.

\bibitem{EMQX4}
E.~Technologies, ``Emqx,'' \url{https://www.emqx.com/en/blog/emqx-vs-mosquitto-2023-mqtt-broker-comparison}, last accessed: \today.

\bibitem{EclipseMosquitto6}
Eclipse, ``Eclipse mosquitto,'' \url{https://www.emqx.com/en/blog/mosquitto-mqtt-broker-pros-cons-tutorial-and-modern-alternatives}, last accessed: \today.

\bibitem{EclipseMosquitto3}
------, ``Eclipse mosquitto,'' \url{https://projects.eclipse.org/projects/iot.mosquitto}, last accessed: \today.

\bibitem{zeromq3}
iMatix, ``Zero mq,'' \url{https://www.hivemq.com/article/mqtt-vs-zeromq-for-iot/}, last accessed: \today.

\bibitem{zeromq2}
------, ``Zero mq,'' \url{http://wiki.zeromq.org/area:faq}, last accessed: \today.

\bibitem{SamZa2}
A.~S. Foundation, ``Apache samza,'' \url{https://samza.incubator.apache.org/learn/documentation/0.7.0/comparisons/storm.html}, last accessed: \today.

\bibitem{kleppmann2019apache}
M.~Kleppmann, ``Apache samza.'' 2019.

\bibitem{EMQX5}
E.~Technologies, ``Emqx,'' \url{https://www.emqx.com/en/blog/emqx-vs-vernemq-2023-mqtt-broker-comparison}, last accessed: \today.

\bibitem{NServiceBus1}
P.~Software, ``Nservicebus,'' \url{https://docs.particular.net/nservicebus/}, last accessed: \today.

\bibitem{NServiceBus4}
------, ``Nservicebus,'' \url{https://docs.particular.net/tutorials/monitoring-setup/}, last accessed: \today.

\bibitem{kestrel2}
Twitter, ``kestrel,'' \url{https://github.com/memcached/memcached/blob/master/doc/protocol.txt}, last accessed: \today.

\bibitem{NATS1}
S.~Communications, ``Nats,'' \url{https://docs.nats.io/}, last accessed: \today.

\bibitem{ayaz2022data}
B.~Ayaz, N.~Slamnik-Krije{\v{s}}torac, and J.~Marquez-Barja, ``Data management platform for smart orchestration of decentralized and heterogeneous vehicular edge networks,'' in \emph{Proceedings of the 2022 ACM Conference on Information Technology for Social Good}, 2022, pp. 118--124.

\bibitem{ibmmq1}
IBM, ``Ibm mq,'' \url{https://cloud.ibm.com/catalog/services/mq}, last accessed: \today.

\bibitem{IronMQibm}
Iron.io, ``Ironmq,'' \url{https://blog.iron.io/ibm-mq-vs-ironmq-pros-cons-and-choosing-an-mq/#4}, last accessed: \today.

\bibitem{ibmmq9}
IBM, ``Ibm mq,'' \url{https://www.ibm.com/}, last accessed: \today.

\bibitem{SQS3}
A.~W.~S. (AWS), ``Amazon sqs,'' \url{https://docs.aws.amazon.com/AWSSimpleQueueService/latest/SQSDeveloperGuide/quotas-messages.html}, last accessed: \today.

\bibitem{MSMQ7}
Microsoft, ``Msmq,'' \url{https://techcommunity.microsoft.com/t5/skype-for-business-blog/troubleshooting-microsoft-message-queuing-issues-on-microsoft/ba-p/619639}, last accessed: \today.

\bibitem{Oracle}
Oracle, ``Oracle message broker,'' \url{https://docs.oracle.com/cd/}, last accessed: \today.

\bibitem{TIBCO1}
TIBCO, ``Tibco enterprise message service™,'' \url{https://docs.tibco.com/pub/ems/}, last accessed: \today.

\bibitem{TIBCO3}
------, ``Tibco enterprise message service™,'' \url{https://support.tibco.com/s/article/Tibco-KnowledgeArticle-Article-22588}, last accessed: \today.

\bibitem{TIBCORendezvous3}
------, ``Tibco rendezvous,'' \url{https://www.tibco.com/products/tibco-cloud-events/pricing-plans}, last accessed: \today.

\bibitem{TIBCORendezvous2}
------, ``Tibco rendezvous,'' \url{https://docs.tibco.com/pub/ems/8.6.0/doc/html/GUID-66774B42-2A5F-4221-864E-3331622E1091.html}, last accessed: \today.

\bibitem{TIBCORendezvous1}
------, ``Tibco rendezvous,'' \url{https://docs.tibco.com/pub/rendezvous/}, last accessed: \today.

\bibitem{AnypointMQ3}
MuleSoft, ``Anypoint mq,'' \url{https://www.mulesoft.com/}, last accessed: \today.

\bibitem{Azure}
Microsoft, ``Azure storage queue,'' \url{https://learn.microsoft.com/en-us/azure/service-bus-messaging/service-bus-azure-and-service-bus-queues-compared-contrasted}, last accessed: \today.

\bibitem{AzureServiceBus5}
------, ``Azure service bus,'' \url{https://learn.microsoft.com/en-us/dotnet/api/azure.messaging.servicebus.}\newline\url{servicebussender.schedulemessageasync?view=azure-dotnet}, last accessed: \today.

\bibitem{AzureServiceBus4}
------, ``Azure service bus,'' \url{https://azure.microsoft.com/en-us/products/service-bus}, last accessed: \today.

\bibitem{SAP3}
S.~AG, ``Sap netweaver process integration,'' \url{https://blogs.sap.com/2014/04/02/message-size-as-source-of-performance-bottleneck/}, last accessed: \today.

\bibitem{Solace3}
C.~Betts, ``Solace,'' \url{https://docs.solace.com/}, last accessed: \today.

\bibitem{Solace1}
------, ``Solace,'' \url{https://www.solace.dev/}, last accessed: \today.

\bibitem{GoogleCloudPub/Sub1}
Google, ``Google cloud pub/sub,'' \url{https://cloudplatform.googleblog.com/2015/04/big-data-cloud-way.html}, last accessed: \today.

\bibitem{AmazonMQ2}
A.~W.~S. (AWS), ``Amazon mq,'' \url{https://docs.aws.amazon.com/amazon-mq/latest/}, last accessed: \today.

\bibitem{IntelMPILibrary2}
Intel, ``Intel mpi library,'' \url{https://texas.gs.shi.com/product/32703496/Intel-MPI-Library-for-Windows}, last accessed: \today.

\bibitem{IntelMPILibrary1}
------, ``Intel mpi library,'' \url{https://www.intel.com/content/www/us/en/developer/articles/technical/improve-performance-and-stability-with-intel-mpi-library-on-infiniband.html}, last accessed: \today.

\bibitem{Kinesis5}
Amazon, ``Kinesis,'' \url{https://repost.aws/knowledge-center/troubleshoot-kinesis-agent-linux}, last accessed: \today.

\bibitem{Kinesis6}
------, ``Kinesis,'' \url{https://docs.aws.amazon.com/streams/latest/}, last accessed: \today.

\bibitem{Kinesis3}
------, ``Kinesis,'' \url{https://docs.aws.amazon.com/kinesisanalytics/latest/dev/error-handling.html}, last accessed: \today.

\bibitem{IronMQ2}
Iron.io, ``Ironmq,'' \url{https://try.iron.io/pricing-worker-monthly/}, last accessed: \today.

\bibitem{IronMQ1}
------, ``Ironmq,'' \url{https://blog.iron.io/apache-kafka-vs-ironmq-whats-best-for-your-business/}, last accessed: \today.

\bibitem{OpenMQmonitor}
Oracle, ``Open message queue, administration guide, release 5.0,'' \url{https://javaee.github.io/glassfish/doc/4.0/mq-admin-guide.pdf}, 2013.

\bibitem{Oracle5}
------, ``Oracle message broker,'' \url{https://docs.oracle.com/cd/E19316-01/820-6424/aerbz/index.html}, last accessed: \today.

\bibitem{Oracle4}
------, ``Oracle message broker,'' \url{https://docs.oracle.com/cd/E19879-01/821-0028/aercs/index.html}, last accessed: \today.

\bibitem{RedHat3}
I.~Red~Hat, ``Red hat amq,'' \url{https://access.redhat.com/documentation/en-us/red_hat_amq/6.1/html/product_introduction/fmbscalable}, last accessed: \today.

\bibitem{brown2020language}
T.~Brown, B.~Mann, N.~Ryder, M.~Subbiah, J.~D. Kaplan, P.~Dhariwal, A.~Neelakantan, P.~Shyam, G.~Sastry, A.~Askell \emph{et~al.}, ``Language models are few-shot learners,'' \emph{Advances in neural information processing systems}, vol.~33, pp. 1877--1901, 2020.

\bibitem{garcia2023we}
F.~Garc{\'\i}a-Pe{\~n}alvo and A.~V{\'a}zquez-Ingelmo, ``What do we mean by genai? a systematic mapping of the evolution, trends, and techniques involved in generative ai,'' 2023.

\bibitem{10198233}
M.~Gupta, C.~Akiri, K.~Aryal, E.~Parker, and L.~Praharaj, ``From chatgpt to threatgpt: Impact of generative ai in cybersecurity and privacy,'' \emph{IEEE Access}, vol.~11, pp. 80\,218--80\,245, 2023.

\bibitem{xu2024large}
S.~Xu, C.~K. Thomas, O.~Hashash, N.~Muralidhar, W.~Saad, and N.~Ramakrishnan, ``Large multi-modal models (lmms) as universal foundation models for ai-native wireless systems,'' \emph{arXiv preprint arXiv:2402.01748}, 2024.

\bibitem{gong2023multimodal}
T.~Gong, C.~Lyu, S.~Zhang, Y.~Wang, M.~Zheng, Q.~Zhao, K.~Liu, W.~Zhang, P.~Luo, and K.~Chen, ``Multimodal-gpt: A vision and language model for dialogue with humans,'' \emph{arXiv preprint arXiv:2305.04790}, 2023.

\bibitem{alayrac2022flamingo}
J.-B. Alayrac, J.~Donahue, P.~Luc, A.~Miech, I.~Barr, Y.~Hasson, K.~Lenc, A.~Mensch, K.~Millican, M.~Reynolds \emph{et~al.}, ``Flamingo: a visual language model for few-shot learning,'' \emph{Advances in Neural Information Processing Systems}, vol.~35, pp. 23\,716--23\,736, 2022.

\bibitem{shen2023hugginggpt}
Y.~Shen, K.~Song, X.~Tan, D.~Li, W.~Lu, and Y.~Zhuang, ``Hugginggpt: Solving ai tasks with chatgpt and its friends in huggingface,'' \emph{arXiv preprint arXiv:2303.17580}, 2023.

\bibitem{huang2023audiogpt}
R.~Huang, M.~Li, D.~Yang, J.~Shi, X.~Chang, Z.~Ye, Y.~Wu, Z.~Hong, J.~Huang, J.~Liu \emph{et~al.}, ``Audiogpt: Understanding and generating speech, music, sound, and talking head,'' \emph{arXiv preprint arXiv:2304.12995}, 2023.

\bibitem{openai2023gpt4}
OpenAI, ``Gpt-4 technical report,'' 2023.

\bibitem{wu2023visual}
C.~Wu, S.~Yin, W.~Qi, X.~Wang, Z.~Tang, and N.~Duan, ``Visual chatgpt: Talking, drawing and editing with visual foundation models,'' \emph{arXiv preprint arXiv:2303.04671}, 2023.

\bibitem{xi2023rise}
Z.~Xi, W.~Chen, X.~Guo, W.~He, Y.~Ding, B.~Hong, M.~Zhang, J.~Wang, S.~Jin, E.~Zhou \emph{et~al.}, ``The rise and potential of large language model based agents: A survey,'' \emph{arXiv preprint arXiv:2309.07864}, 2023.

\bibitem{wang2023survey}
L.~Wang, C.~Ma, X.~Feng, Z.~Zhang, H.~Yang, J.~Zhang, Z.~Chen, J.~Tang, X.~Chen, Y.~Lin \emph{et~al.}, ``A survey on large language model based autonomous agents,'' \emph{arXiv preprint arXiv:2308.11432}, 2023.

\bibitem{wei2022chain}
J.~Wei, X.~Wang, D.~Schuurmans, M.~Bosma, F.~Xia, E.~Chi, Q.~V. Le, D.~Zhou \emph{et~al.}, ``Chain-of-thought prompting elicits reasoning in large language models,'' \emph{Advances in Neural Information Processing Systems}, vol.~35, pp. 24\,824--24\,837, 2022.

\bibitem{yao2022react}
S.~Yao, J.~Zhao, D.~Yu, N.~Du, I.~Shafran, K.~Narasimhan, and Y.~Cao, ``React: Synergizing reasoning and acting in language models,'' \emph{arXiv preprint arXiv:2210.03629}, 2022.

\bibitem{shinn2023reflexion}
N.~Shinn, F.~Cassano, A.~Gopinath, K.~R. Narasimhan, and S.~Yao, ``Reflexion: Language agents with verbal reinforcement learning,'' in \emph{Thirty-seventh Conference on Neural Information Processing Systems}, 2023.

\bibitem{chatgpt2023}
openai, ``Chatgpt,'' \url{https://chat.openai.com/}, last accessed: \today.

\bibitem{shah2023lm}
D.~Shah, B.~Osi{\'n}ski, S.~Levine \emph{et~al.}, ``Lm-nav: Robotic navigation with large pre-trained models of language, vision, and action,'' in \emph{Conference on Robot Learning}.\hskip 1em plus 0.5em minus 0.4em\relax PMLR, 2023, pp. 492--504.

\bibitem{mu2023embodiedgpt}
Y.~Mu, Q.~Zhang, M.~Hu, W.~Wang, M.~Ding, J.~Jin, B.~Wang, J.~Dai, Y.~Qiao, and P.~Luo, ``Embodiedgpt: Vision-language pre-training via embodied chain of thought,'' \emph{arXiv preprint arXiv:2305.15021}, 2023.

\bibitem{qian2023creator}
C.~Qian, C.~Han, Y.~R. Fung, Y.~Qin, Z.~Liu, and H.~Ji, ``Creator: Disentangling abstract and concrete reasonings of large language models through tool creation,'' \emph{arXiv preprint arXiv:2305.14318}, 2023.

\bibitem{chen2023teaching}
X.~Chen, M.~Lin, N.~Sch{\"a}rli, and D.~Zhou, ``Teaching large language models to self-debug,'' \emph{arXiv preprint arXiv:2304.05128}, 2023.

\bibitem{park2019wireless}
J.~Park, S.~Samarakoon, M.~Bennis, and M.~Debbah, ``Wireless network intelligence at the edge,'' \emph{Proceedings of the IEEE}, vol. 107, no.~11, pp. 2204--2239, 2019.

\bibitem{loven2019edgeai}
L.~Lov{\'e}n, T.~Lepp{\"a}nen, E.~Peltonen, J.~Partala, E.~Harjula, P.~Porambage, M.~Ylianttila, and J.~Riekki, ``Edgeai: A vision for distributed, edge-native artificial intelligence in future 6g networks,'' \emph{6G Wireless Summit, March 24-26, 2019 Levi, Finland}, 2019.

\bibitem{deng2020edge}
S.~Deng, H.~Zhao, W.~Fang, J.~Yin, S.~Dustdar, and A.~Y. Zomaya, ``Edge intelligence: The confluence of edge computing and artificial intelligence,'' \emph{IEEE Internet of Things Journal}, vol.~7, no.~8, pp. 7457--7469, 2020.

\bibitem{dhoni2023exploring}
P.~Dhoni, ``Exploring the synergy between generative ai, data and analytics in the modern age,'' \emph{Authorea Preprints}, 2023.

\bibitem{chiu2023impact}
T.~K. Chiu, ``The impact of generative ai (genai) on practices, policies and research direction in education: A case of chatgpt and midjourney,'' \emph{Interactive Learning Environments}, pp. 1--17, 2023.

\bibitem{openai2023}
OpenAI, ``Gpt-4 system card,'' \url{https://cdn.openai.com/papers/gpt-4-system-card.pdf}, last accessed: \today.

\bibitem{celik2024dawn}
A.~Celik and A.~M. Eltawil, ``At the dawn of generative ai era: A tutorial-cum-survey on new frontiers in 6g wireless intelligence,'' \emph{IEEE Open Journal of the Communications Society}, 2024.

\bibitem{jin2023large}
M.~Jin, Q.~Wen, Y.~Liang, C.~Zhang, S.~Xue, X.~Wang, J.~Zhang, Y.~Wang, H.~Chen, X.~Li \emph{et~al.}, ``Large models for time series and spatio-temporal data: A survey and outlook,'' \emph{arXiv preprint arXiv:2310.10196}, 2023.

\bibitem{zhang2024large}
X.~Zhang, R.~R. Chowdhury, R.~K. Gupta, and J.~Shang, ``Large language models for time series: A survey,'' \emph{arXiv preprint arXiv:2402.01801}, 2024.

\bibitem{rawte2023survey}
V.~Rawte, A.~Sheth, and A.~Das, ``A survey of hallucination in large foundation models,'' \emph{arXiv preprint arXiv:2309.05922}, 2023.

\bibitem{gupta2023chatgpt}
M.~Gupta, C.~Akiri, K.~Aryal, E.~Parker, and L.~Praharaj, ``From chatgpt to threatgpt: Impact of generative ai in cybersecurity and privacy,'' \emph{IEEE Access}, 2023.

\bibitem{bang2023examination}
J.~Bang, B.-T. Lee, and P.~Park, ``Examination of ethical principles for llm-based recommendations in conversational ai,'' in \emph{2023 International Conference on Platform Technology and Service (PlatCon)}.\hskip 1em plus 0.5em minus 0.4em\relax IEEE, 2023, pp. 109--113.

\bibitem{naik2017choice}
N.~Naik, ``Choice of effective messaging protocols for iot systems: Mqtt, coap, amqp and http,'' in \emph{2017 IEEE international systems engineering symposium (ISSE)}.\hskip 1em plus 0.5em minus 0.4em\relax IEEE, 2017, pp. 1--7.

\bibitem{min2023recent}
B.~Min, H.~Ross, E.~Sulem, A.~P.~B. Veyseh, T.~H. Nguyen, O.~Sainz, E.~Agirre, I.~Heintz, and D.~Roth, ``Recent advances in natural language processing via large pre-trained language models: A survey,'' \emph{ACM Computing Surveys}, vol.~56, no.~2, pp. 1--40, 2023.

\bibitem{xia2023generative}
L.~Xia, Y.~Sun, C.~Liang, L.~Zhang, M.~A. Imran, and D.~Niyato, ``Generative ai for semantic communication: Architecture, challenges, and outlook,'' \emph{arXiv preprint arXiv:2308.15483}, 2023.

\bibitem{jiang2023semantic}
P.~Jiang, C.-K. Wen, X.~Yi, X.~Li, S.~Jin, and J.~Zhang, ``Semantic communications using foundation models: Design approaches and open issues,'' \emph{arXiv preprint arXiv:2309.13315}, 2023.

\bibitem{donta2023governance}
P.~K. Donta, B.~Sedlak, V.~Casamayor~Pujol, and S.~Dustdar, ``Governance and sustainability of distributed continuum systems: a big data approach,'' \emph{Journal of Big Data}, vol.~10, no.~1, pp. 1--31, 2023.

\bibitem{oniani2023military}
D.~Oniani, J.~Hilsman, Y.~Peng, R.~K. Poropatich, C.~Pamplin, L.~Legault, Y.~Wang \emph{et~al.}, ``From military to healthcare: Adopting and expanding ethical principles for generative artificial intelligence,'' \emph{arXiv preprint arXiv:2308.02448}, 2023.

\bibitem{zhang2023mlcopilot}
L.~Zhang, Y.~Zhang, K.~Ren, D.~Li, and Y.~Yang, ``Mlcopilot: Unleashing the power of large language models in solving machine learning tasks,'' \emph{arXiv preprint arXiv:2304.14979}, 2023.

\bibitem{shen2023large}
Y.~Shen, J.~Shao, X.~Zhang, Z.~Lin, H.~Pan, D.~Li, J.~Zhang, and K.~B. Letaief, ``Large language models empowered autonomous edge ai for connected intelligence,'' \emph{arXiv preprint arXiv:2307.02779}, 2023.

\bibitem{li2023colossal}
S.~Li, H.~Liu, Z.~Bian, J.~Fang, H.~Huang, Y.~Liu, B.~Wang, and Y.~You, ``Colossal-ai: A unified deep learning system for large-scale parallel training,'' in \emph{Proceedings of the 52nd International Conference on Parallel Processing}, 2023, pp. 766--775.

\bibitem{li2021sequence}
S.~Li, F.~Xue, C.~Baranwal, Y.~Li, and Y.~You, ``Sequence parallelism: Long sequence training from system perspective,'' \emph{arXiv preprint arXiv:2105.13120}, 2021.

\bibitem{korthikanti2023reducing}
V.~A. Korthikanti, J.~Casper, S.~Lym, L.~McAfee, M.~Andersch, M.~Shoeybi, and B.~Catanzaro, ``Reducing activation recomputation in large transformer models,'' \emph{Proceedings of Machine Learning and Systems}, vol.~5, 2023.

\bibitem{radford2018improving}
A.~Radford, K.~Narasimhan, T.~Salimans, I.~Sutskever \emph{et~al.}, ``Improving language understanding by generative pre-training,'' 2018.

\bibitem{micikevicius2018mixed}
P.~Micikevicius, S.~Narang, J.~Alben, G.~Diamos, E.~Elsen, D.~Garcia, B.~Ginsburg, M.~Houston, O.~Kuchaiev, G.~Venkatesh, and H.~Wu, ``Mixed precision training,'' 2018.

\bibitem{jiang2022model}
Y.~Jiang, S.~Wang, V.~Valls, B.~J. Ko, W.-H. Lee, K.~K. Leung, and L.~Tassiulas, ``Model pruning enables efficient federated learning on edge devices,'' \emph{IEEE Transactions on Neural Networks and Learning Systems}, 2022.

\bibitem{zhu2023survey}
X.~Zhu, J.~Li, Y.~Liu, C.~Ma, and W.~Wang, ``A survey on model compression for large language models,'' \emph{arXiv preprint arXiv:2308.07633}, 2023.

\bibitem{polino2018model}
A.~Polino, R.~Pascanu, and D.~Alistarh, ``Model compression via distillation and quantization,'' \emph{arXiv preprint arXiv:1802.05668}, 2018.

\bibitem{hsu2022language}
Y.-C. Hsu, T.~Hua, S.~Chang, Q.~Lou, Y.~Shen, and H.~Jin, ``Language model compression with weighted low-rank factorization,'' \emph{arXiv preprint arXiv:2207.00112}, 2022.

\bibitem{kreuzberger2023machine}
D.~Kreuzberger, N.~Kühl, and S.~Hirschl, ``Machine learning operations (mlops): Overview, definition, and architecture,'' \emph{IEEE Access}, vol.~11, pp. 31\,866--31\,879, 2023.

\end{thebibliography}

\end{document}